\documentclass[prc,showpacs,preprintnumbers,amsmath,amssymb]{revtex4}
\usepackage{graphicx}

\begin{document}
\title{Photoproduction of K\mbox{\boldmath$\Lambda$} on the proton}
\author{D. Skoupil and P. Byd\v{z}ovsk\'y}
\affiliation{Nuclear Physics Institute, ASCR, \v{R}e\v{z}/Prague, 
Czech Republic}

\date{\today }

\begin{abstract}
Kaon photoproduction on the proton is studied in the resonance region 
using an isobar model. The higher-spin nucleon (3/2 and 5/2) and hyperon (3/2) 
resonances were included in the model utilizing the consistent formalism 
by Pascalutsa, and they were found to play an important role in data 
description. The spin-1/2 and spin-3/2 hyperon resonances in combination with 
the Born terms contribute significantly to the background part of the amplitude. 
Various forms of the hadron form factor were considered in the construction, 
and the dipole and multidipole forms were selected as those most suitable  
for the data description. Model parameters were fitted to new experimental 
data from CLAS, LEPS, and GRAAL collaborations and two versions of the model, 
BS1 and BS2, were chosen. Both models provide a good overall description of 
the data for the center-of-mass energies from the threshold up to 2.4 GeV. 
Predicted cross sections of the models at very small kaon angles 
  being consistent with results of the Saclay-Lyon model indicate that 
  the models could be also successful in predicting the hypernucleus 
  production cross sections. 
Although kaon photoproduction takes place in the third-resonance region with 
many resonant states, the total number of included resonances, 
15 and 16, is quite moderate and it is comparable with numbers of resonances 
in other models.
The set of chosen nucleon resonances overlaps well with 
the set of the most probable contributing states determined in the Bayesian analysis with the Regge-plus-resonance model. Particularly, we confirm that the missing 
resonances $P_{13}(1900)$ and $D_{13}(1875)$ do play an important role in 
the description of data. However, the spin-1/2 state $P_{11}(1880)$ included 
in the Bayesian analysis was replaced in our analysis with the near-mass 
spin-5/2 state $N^*(1860)$, recently considered by the Particle Data Group.
\end{abstract}


\maketitle
%
%
\section{Introduction}
\label{Introduction}
The investigation of kaon-hyperon photo- and electroproduction from nucleons 
in the nucleon resonance region provides important information about the baryon 
resonance spectrum and interactions in hyperon-nucleon systems arising from 
QCD. Besides studying the reaction mechanism, one can learn more about the existence and properties of the ``missing'' resonances that are predicted by the quark models~\cite{CapRob00,Bonn} but that are weakly coupled to the $\pi$N final state and therefore are not seen in the pion production or $\pi$N scattering processes. A correct description of the elementary $\Lambda$-production process is also important for getting reliable predictions of the excitation spectra for production of $\Lambda$ hypernuclei~\cite{HProd,ByMa}.

Numerous theoretical studies of the hyperon production have been performed 
over the past decades. The analyses before 2004, however, suffered from 
a lack of high-quality experimental data, see \emph{e.g.}, 
Refs.~\cite{AS,SL,SLA,Jan01A,Jan01B,Jan02,WJC,models} and references 
therein.   
The situation changed significantly after new high-duty-factor accelerators, 
providing good quality high-current polarized continuous beams, were 
constructed in Jefferson Lab (CEBAF) and Bonn University (ELSA). Number 
of good quality data, especially from the CLAS~\cite{CLAS05,CLAS10}, 
LEPS~\cite{LEPS}, GRAAL~\cite{GRAAL}, and SAPHIR~\cite{SAPHIR}  
collaborations, rose by more than a factor of 10 which revived interest 
in modeling the 
process~\cite{Giessen,JDiaz,Anisovich,Borasoy,GentRPR07,MartSu,Mart,Maxwell}.
Now various response functions are accessible and measured with a good level 
of precision in the energy region from the threshold up to 2.8 GeV, which 
allows us to perform more rigorous tests of theoretical models and  
improve our understanding of the elementary process. 

The models of $\gamma\rm p \longrightarrow  K^+\Lambda$ that are in 
a close connection with QCD are based on quark degrees of 
freedom~\cite{ZpL95,LLP95,FHZ91}. These quark models need a relatively 
small number of parameters and assume explicitly a spatial-extended 
structure of the baryons which was found to be important for a reasonable 
description of the photoproduction data~\cite{LLP95}. Contributions 
of baryon resonances in the intermediate state then arise naturally from 
effects of excited states of the quark system. Alternative approaches  
to description of the production process at low energies assume hadrons as 
appropriate effective degrees of freedom. Calculations grounded in 
an effective Lagrangian containing interacting meson, baryon and 
electromagnetic fields provide us with a valuable tool for analysis 
of experimental data. As there is no explicit connection to QCD, 
the number of parameters in the models is related to the number 
of resonances included in the calculations and is, therefore, relatively 
large for the kaon production~\cite{SL,SLA,Jan01A,Jan01B,Jan02,MartSu}. 
The short-range physics manifesting itself via a spatial structure of 
hadrons can be simulated by a form factor introduced in the interaction 
vertex. This, however, brings another ambiguity into the model: the forms and parameters of the form factors, which have to be fixed in 
a data analysis.

In some models the concept of chiral symmetry is utilized to include 
the pseudoscalar mesons as the Goldstone bosons in the chiral quark 
model~\cite{ZpL95} or to build up a chiral effective meson-baryon 
Lagrangian in the gauge-invariant chiral unitary model~\cite{Borasoy}. 
Attempts were also made to calculate the kaon-hyperon photoproduction 
processes in the threshold region in the framework of the chiral 
perturbation theory~\cite{Chpt}.
 
In the hadrodynamical approach, the production channels are coupled 
by the meson-baryon interaction in the final state and should 
be, therefore, treated simultaneously to maintain unitarity. 
In the coupled-channel approaches~\cite{Giessen,JDiaz,Anisovich,Borasoy}, 
the rescattering effects in the meson-baryon final-state system are 
included, but the models face the problem of missing experimental 
information on some transition amplitudes, \emph{e.g.}, 
${\rm K^+}\Lambda \longrightarrow  {\rm K^+}\Lambda$.   
Considerable simplification originates from neglecting the rescattering 
effects in the formalism, assuming that their influence on the results 
is included to some extent by means of effective values of the coupling 
constants fitted to experimental data. 
This simplifying assumption was adopted in many single-channel isobar 
models, \emph{e.g.}, Saclay-Lyon (SL)~\cite{SL,SLA}, Kaon-MAID (KM)~\cite{KM}, 
and Gent-Isobar~\cite{Jan01A,Jan01B,Jan02}. Unitarity corrections in the 
single-channel approach can be included by energy-dependent widths 
in the resonance propagators~\cite{KM}. Since the early work of 
Thom~\cite{Thom}, the isobar models were among the first models 
capable of describing the kaon photoproduction in the resonance region.

The kaon production takes place in the third-resonance region, with 
many possible nucleon and hyperon higher-spin states coupling to 
the kaon-hyperon channels. Therefore, the contributions of higher-spin 
baryon resonances are particularly important in the isobar models. 
The theoretical description of the interacting baryon fields with a spin 
higher than 1/2  causes problems due to the presence of nonphysical 
lower-spin components in the Rarita-Schwinger field~\cite{Benm}. 
Some prescriptions for the propagator and vertexes had to be adopted 
to handle the higher-spin problem; see, \emph{e.g.}, Refs.~\cite{SL,SLA} for 
the case of spin-3/2 nucleon resonances. 
This prescription, however, requires fixing additional free parameters 
in the Lagrangian, the off-shell parameters~\cite{SLA,Jan01A,Jan01B,Jan02}. 
Moreover, the prescription used in Ref.~\cite{SL} did not allow inclusion of 
the hyperon resonances with spin 3/2 due to the terms in the propagator 
diverging for the $u$-channel exchanges. These divergences were removed in 
Ref.~\cite{SLA} by considering the correct propagator for massive spin-3/2 
particle~\cite{Benm}, which, however, contains the spin-1/2 contribution.
These problems were removed by 
Pascalutsa who formulated a consistent theory for massive spin-3/2 fields 
requiring invariance of the interactions under the local gauge transformation 
of the Rarita-Schwinger field~\cite{Pasc}. This formalism was recently 
generalized to arbitrary spin by the Gent group~\cite{Gent_spin}.   

Description of the kaon-hyperon photo- and electroproduction from the 
threshold up to energies rather above the resonance region 
($E_\gamma \approx 16$~GeV) is possible with the Regge-plus-resonance model (RPR)
constructed by the Gent group~\cite{GentRPR07,RPR11}. 
This hybrid model combines the Regge model~\cite{Guidal}, appropriate 
for description above the resonance region ($E_\gamma > 3$~GeV), with 
the isobar model eligible for description in the resonance region. 
The Regge-based part of the amplitude, which is a smooth function of 
energy, constitutes the main contribution to the background 
in the resonance region. The resonance part of the amplitude 
is modeled by the \emph{s}-channel exchanges of nucleon resonances, 
with strong hadron form factors ensuring that these resonant 
contributions vanish above the resonance region where the Regge 
part dominates. 
This concept significantly reduces the number of background parameters 
in comparison with an isobar model, and removes the necessity to introduce 
the hadron form factors in the background to reduce too large 
contributions from  the Born terms~\cite{Jan01A,Jan01B,KM}.

In this work, we have constructed a new isobar model for photoproduction 
of K$\Lambda$; however, most of the  presented formulas are valid also 
for K$\Lambda$ electroproduction. We have used the new consistent formalism 
for the description of the higher-spin baryon resonances by 
Pascalutsa~\cite{Pasc,Gent_spin}, allowing us to include the hyperon resonances 
with spin 3/2. 
We also paid attention to properties of the model at very forward 
kaon-angle production which is relevant to calculations of the cross 
sections in the hypernucleus photoproduction~\cite{HProd,ByMa}. 
This article is organized as follows: In Sec. II we present important 
ingredients of our model. The basic formalism is given in Sec. III. 
The method of fitting free model parameters to experimental data is 
described in Sec. IV. Discussion of obtained results and conclusions are 
presented in Secs. V and VI. Contributions to the invariant amplitude 
from the Feynman diagrams are given in the Appendix.
%
%
\section{Single-channel isobar model}
\label{isobarmodel}
In this section, we give the main features of the theoretical 
framework used in our approach. For other details we refer the reader 
to Refs.~\cite{SL,SF} and references therein. Here we investigate the 
K$\Lambda$ photoproduction on the proton at center-of-mass (c.m.) energies $\le 2.5$~GeV, 
but the presented formulas can be used also for electroproduction. 

\begin{figure}[h]
\includegraphics[width=10cm]{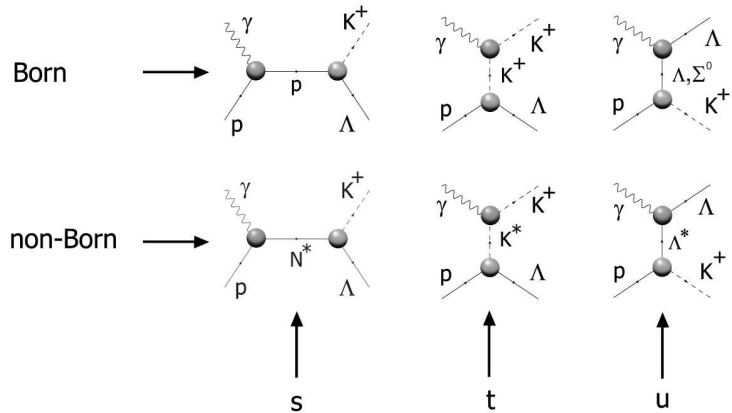}
\caption{The tree-level contributions to the $p(\gamma,K^+)\Lambda$ amplitude 
are shown. The Born terms with a ground-state hadron exchanges and 
the non-Born terms with nucleon-, kaon- and hyperon-resonance exchanges 
are shown in the upper and lower rows, respectively.}
\label{fig:stu}
\end{figure}

In the isobar model, the amplitude is constructed from an effective 
meson-baryon Lagrangian as a sum of the tree-level Feynman diagrams 
representing the {\em s}-, {\em t}-, and {\em u}-channel exchanges 
of the ground-state hadrons (the Born terms) and various resonances 
(the non-Born terms), see Fig.~\ref{fig:stu}. The higher-order 
contributions, that account for, \emph{e.g.}, the rescattering effects, 
are neglected. Only the exchanges 
of nucleon resonances in the \emph{s}-channel make a resonant structure 
in the observables. The other diagrams contribute to the background part 
of the amplitude as the corresponding poles are far from the physical region.

Since there exists no dominant resonance in photoproduction of kaons, 
unlike in $\pi$ or $\eta$ photoproduction, one has to take into account 
{\em a priori} more than 20 resonances with the mass $\leq 2\,\mbox{GeV}$ 
(see Tab.~\ref{TableAllRes}). 
This leads to a huge number of possible resonance configurations that should 
be investigated~\cite{AS,SL,RPR11}, still resulting in a large number 
of models that describe the data quite well (with a small $\chi^2$). 
To reduce this large number of models one imposes constraints to 
acceptable values of the K$\Lambda$N and K$\Sigma$N coupling constants 
relating them to the well known $\pi$NN value by means of the SU(3) 
symmetry~\cite{AS,SL,Jan01A}. 

One of the characteristic features of the $p(\gamma, K^+)\Lambda$ process 
described by an isobar model is a too large contribution of the Born terms 
to the cross sections which largely overpredicts the data. 
To get a realistic description of the cross sections and the other observables  
allowing analysis of the resonant content of the amplitude, the nonphysically 
large strengths of the Born terms have to be reduced. This can be achieved 
by the introduction of form factors into the strong vertexes (hadron form 
factors)~\cite{KM} or by exchanges of several hyperon resonances~\cite{SL} or 
by a combination of both methods~\cite{Jan01A,Jan01B}. 
In our model we combine both methods. Needless to say, the choice 
of the method strongly affects the dynamics of the model. 
Note that this problem is not present in the Regge-plus-resonance 
model~\cite{GentRPR07,RPR11}.

Another ambiguity in construction of the gauge-invariant Lagrangian arises 
from a coupling in the K$\Lambda$N vertex which can be either pseudoscalar- 
or pseudovector-like~\cite{PSPV}. 
Whereas the former makes the total contribution of the Born terms gauge 
invariant the use of the latter requires introducing the contact term 
even with no form factors inserted. 
The role of these couplings was investigated in the threshold region~\cite{Mart} 
and it was concluded that both couplings can describe the K$\Lambda$  photoproduction 
data very well.  In this work we have used the pseudoscalar coupling as 
in the most of isobar models.

To ensure a regularity of the tree-level invariant amplitude in 
the physical region the poles corresponding to the resonances are shifted 
into the complex plane, $m_R\to m_R-\texttt{i}\Gamma_R /2$, introducing 
the decay width $\Gamma_R$ which accounts for a  finite lifetime of 
the resonant state. Then the Feynman propagator can be written as 
\begin{equation}
\frac{1}{\not\! q-m_R+\texttt{i}\Gamma_R /2} = 
\frac{\not\! q+m_R-\texttt{i}\Gamma_R /2}{q^2-m^2_R+\texttt{i}m_R\Gamma_R+\Gamma_R^2 /4}
\end{equation}
and the following approximations are assumed in various isobar models
\begin{equation}
\frac{\not\! q+m_R}{q^2-m^2_R+\texttt{i}m_R\Gamma_R}
\label{SLprop}
\end{equation}
in the Saclay-Lyon and Gent models or
\begin{equation}
\frac{\not\! q+m_R-\texttt{i}\Gamma_R /2}{q^2-m^2_R+\texttt{i}m_R\Gamma_R}
\end{equation}
in the Kaon-MAID model and in Ref.~\cite{SF}.
In the tree-level approximation, the decay widths can mimic to some extent 
a dressing of the propagator. In most of the isobar models the widths  
are assumed as constant parameters, and the Breit-Wigner values suggested 
in the Particle Data Tables are used. In order to approximately account 
for unitarity corrections in the single-channel approach, the energy-dependent
widths for the nucleon resonances were used in the KM model. 
The energy dependence of $\Gamma_R$ is given by a possibility of resonance 
decay into various opened channels. In this work we 
use the approximation (\ref{SLprop}) with constant decay widths.
\begin{table}[h!]
\hspace{.0cm}
\begin{tabular}{lccccrr}
\hline
& ~Nickname~&~Particle~     & ~Mass~      & ~Width~     & $J^\pi$   & Status\\ 
& & &~[MeV]~&~[MeV]~& & \\ \hline
& ${\rm K}^*$& $K^*(892)$    & 891.66      & 50.8        & $1^-$     &    \\
& K1        & $K_1(1270)$    & 1272        & 90          & $1^+$     &    \\
& N1        & $P_{11}(1440)$ & 1430        & 350         & $1/2^+$   & ****\\
& N3        & $S_{11}(1535)$ & 1535        & 150         & $1/2^-$   & ****\\
& N4        & $S_{11}(1650)$ & 1655        & 150         & $1/2^-$   & ****\\
& N8        & $D_{15}(1675)$ & 1675        & 150         & $5/2^-$   & ****\\
& N9        & $F_{15}(1680)$ & 1685        & 130         & $5/2^+$   & ****\\
& N5        & $D_{13}(1700)$ & 1700        & 150         & $3/2^-$   & ***\\
& N6        & $P_{11}(1710)$ & 1710        & 100         & $1/2^+$   & ***\\
& N7        & $P_{13}(1720)$ & 1720        & 270         & $3/2^+$   & ****\\
& P5        & $F_{15}(1860)$ & 1860        & 270         & $5/2^+$   & **\\
& P1        & $P_{11}(1880)$ & 1870        & 235         & $1/2^+$   & **\\
& P4        & $D_{13}(1875)$ & 1875        & 220         & $3/2^-$   & ***\\
& P2        & $P_{13}(1900)$ & 1900        & 500         & $3/2^+$   & ***\\
& P3        & $F_{15}(2000)$ & 2050        & 198         & $5/2^+$   & **\\
& L1        & $\Lambda(1405)$& 1405        & 50          & $1/2^-$   & ****\\
& L2        & $\Lambda(1600)$& 1600        & 150         & $1/2^+$   & ***\\
& L3        & $\Lambda(1670)$& 1670        & 35          & $1/2^-$   & ****\\
& L4        & $\Lambda(1800)$& 1800        & 300         & $1/2^-$   & ***\\
& L5        & $\Lambda(1810)$& 1810        & 150         & $1/2^+$   & ***\\
& L6        & $\Lambda(1520)$& 1519.54     & 15.6        & $3/2^-$   & ****\\
& L7        & $\Lambda(1690)$& 1690        & 60          & $3/2^-$   & ****\\
& L8        & $\Lambda(1890)$& 1890        & 100         & $3/2^+$   & ****\\
& S1        & $\Sigma(1660)$ & 1660        & 100         & $1/2^+$   & ***\\
& S2        & $\Sigma(1750)$ & 1750        & 90          & $1/2^-$   & ***\\
& S3        & $\Sigma(1670)$ & 1670        & 60          & $3/2^-$   & ****\\
& S4        & $\Sigma(1940)$ & 1940        & 220         & $3/2^-$   & ***\\
\hline
\end{tabular}
\caption{Meson and baryon resonances which can be included in a description of 
the $p(\gamma, K^+)\Lambda$ process. For each resonance, the mass, width, spin, 
parity, and status are shown. Entries are from Particle Data Tables 2014~\cite{PDG} 
except for the P2 width which was taken from the Bayesian analysis of the Gent 
group.} 
\label{TableAllRes}
\end{table}

\subsection{Resonances with spin 3/2 and 5/2}

The Rarita-Schwinger (R-S) description of high-spin fermion fields includes 
nonphysical degrees of freedom connected with their lower-spin content. 
If the R-S field is off its mass shell, the nonphysical parts may participate 
in the interaction, which is then called ``inconsistent''.
Almost two decades ago, Pascalutsa proposed a new consistent interaction 
theory for massive spin-3/2 fields~\cite{Pasc}, where the interaction is 
mediated by the spin-3/2 modes only. The consistency of the theory 
is ensured by the invariance of the spin-3/2 interaction vertexes under 
the local $U(1)$ gauge transformation of the R-S field. This scheme 
was generalized to arbitrary high spin by the Gent group~\cite{Gent_spin}
and is used in this work. 	

The R-S propagator of the spin-3/2 field in terms of 
the spin-projection operators is~\cite{Timm}
\begin{equation}
S_{\mu\nu}(q)=\frac{\not\! q+m_R}{q^{2}-m_R^{2}+\texttt{i}m_R\Gamma_R}\,P_{\mu\nu}^{(3/2)}  
-\frac{2}{3m_R^{2}}(\not\! q+m_R)\,P_{22,\mu\nu}^{(1/2)} 
+\frac{1}{m_R\sqrt{3}}\left(P_{12,\mu\nu}^{(1/2)}+P_{21,\mu\nu}^{(1/2)}\right),
\label{32propagator}
\end{equation}
where $P_{\mu\nu}^{(3/2)}$ projects on the spin-3/2 states
\begin{equation}
P_{\mu\nu}^{(3/2)}=g_{\mu\nu}-\frac{1}{3}\gamma_{\mu}\gamma_{\nu}-\frac{\not\!q q_{\nu}\gamma_{\mu}+q_{\mu}\gamma_{\nu}\not\!q}{3q^{2}},
\label{eq:32projector}
\end{equation}
and  $P_{12,\mu\nu}^{(1/2)}$, $P_{21,\mu\nu}^{(1/2)}$, and 
$P_{22,\mu\nu}^{(1/2)}$ project on the spin-1/2 sector
\begin{equation}
P_{22,\mu\nu}^{(1/2)}=\frac{q_{\mu}q_{\nu}}{q^{2}},\,\,\,\, 
P_{12,\mu\nu}^{(1/2)}=\frac{q^{\rho}q_{\nu}\sigma_{\mu\rho}}{\sqrt{3}q^{2}}, \,\,\,\, 
P_{21,\mu\nu}^{(1/2)}=\frac{q_{\mu}q^{\rho}\sigma_{\rho\nu}}{\sqrt{3}q^{2}},
\label{P12}
\end{equation}
where $\sigma_{\rho\nu}=\frac{\texttt{i}}{2}[\gamma_{\rho},\gamma_{\nu}]$.

The gauge invariance of the strong, $K(p_K)\,\Lambda \,N^*(q)$, 
and electromagnetic, $N^*(q)\,p\,\gamma (k)$, couplings~\cite{Pasc} generates 
the transverse interaction vertexes 
\begin{equation}
V^S_\mu (K\Lambda N^*) = \frac{f}{m_Km_R}\,
\epsilon_{\lambda\mu\alpha\beta}\,\gamma_5\,\gamma^\alpha\,q^\lambda\,
p_K^\beta\, ,
\label{strongV}
\end{equation}
and 
\begin{equation}
V^{EM}_\nu(N^*p\gamma) = \frac{\texttt{i}\,\gamma_5}{m_R(m_R+m_p)}\,q^\tau 
\left[ \; g_1 F_{\tau\nu} + 
g_2\,(\,\gamma_\tau \gamma^\sigma F_{\nu\sigma} - 
\gamma_\nu \gamma^\sigma F_{\tau\sigma} )\, \right]\, , 
\label{EMV}
\end{equation}
where $F_{\mu\nu}=k_\mu\varepsilon_\nu - 
\varepsilon_\mu k_\nu$, $\epsilon_{0123}=1$, and 
\begin{equation}
V^S_\mu\,q^\mu\,=\,V^{EM}_\nu\,q^\nu\,=\,0\;.
\label{transversV}
\end{equation}
Then it is obvious from Eqs.~(\ref{32propagator}) and (\ref{P12})  
that this property removes all non physical contributions of the spin-1/2 
sector to the invariant amplitude. Moreover, one sees in 
Eq.~(\ref{eq:32projector}) that the pole term in $P_{\mu\nu}^{(3/2)}$ also 
vanishes which makes it possible to include into the model the hyperon 
exchanges with the spin 3/2 in the {\em u}-channel (see Subsec. C below).

In general, for arbitrary high spin $n+1/2$ ($n$= 1, 2,..), the transversality 
of the interaction vertexes prevents the momentum-dependent terms 
in the propagator from contributing, allowing us to write the R-S propagator in 
the consistent theory only by means of the projection operator 
onto the pure spin-$(n+1/2)$ state~\cite{Gent_spin}:
\begin{equation}
S_{\mu_1\cdot\cdot\cdot\mu_n,\nu_1\cdot\cdot\cdot\nu_n}(q)\to 
\frac{\not\! q +m_R}{q^2-m_R^2+\texttt{i}m_R\Gamma_R} 
{P}^{(n+1/2)}_{\mu_1\cdot\cdot\cdot\mu_n,\nu_1\cdot\cdot\cdot\nu_n}(q).
\end{equation}

The gauge invariance of the interaction results also in a relatively 
high-power momentum dependence in the invariant amplitude, which rises 
with rising spin of the R-S field as $\sim q^{2n}$~\cite{Gent_spin}.
For the spin-3/2 field it is apparent from Eqs.~(\ref{strongV}) and 
(\ref{EMV}) that the momentum dependence is $\sim q^\lambda q^\tau$, 
see also (\ref{eq:N32ampl}) for the $s$-channel invariant amplitude. 
In the case of spin 5/2, the invariant amplitude can be schematically 
written as
\begin{equation}
\mathbb{M}_{NBs}^{N^*(5/2)} \sim q^4 \frac{\not\! q +m_R}{q^2-m_R^2+\texttt{i}m_R\Gamma_R} 
{\cal P}^{(5/2)}_{\mu\nu ,\lambda\rho}(q)\,{\cal O}^{\mu\nu ,\lambda\rho}_{5/2},
\end{equation}
where ${\cal P}^{(5/2)}_{\mu\nu ,\lambda\rho}(q)$ projects onto the spin-5/2 
state~\cite{Gent_spin} and ${\cal O}^{\mu\nu ,\lambda\rho}_{5/2}$ stands for 
the remaining structure in the strong and electromagnetic vertexes, 
see (\ref{eq:N52ampl}). 

This strong momentum dependence from derivatives in the gauge-invariant 
vertexes regularizes the amplitude, but it also causes nonphysical structures 
in the energy dependence of the cross section, which needs to be cut off 
especially above the resonance region. Therefore, the hadron form factors 
with a higher, spin-dependent energy power in the denominator and with 
relatively small values of the cut-off parameter in comparison with standard 
hadron form factors are used in the RPR model~\cite{Gent_spin,RPR11}.  
We have, therefore, also carefully investigated this property in our 
isobar approach considering various forms of the hadron form factor, 
see Subsec. D below.

Note that after the substitution $\sqrt{s}\to m_R$ the propagator used in 
the SL model~\cite{SL} equals that in Eq.~(\ref{32propagator}). 
The interaction Lagrangians in SL, constructed as the most general 
form invariant under the so-called point transformation~\cite{SLA}, 
lead in general to an inconsistent description. 
Moreover, this point-transformation 
invariance adds three more free parameters, the off-shell parameters, 
to each spin-3/2 resonance~\cite{SLA,Jan01A,Jan01B,Jan02}. Using the consistent 
formalism in our approach we have avoided this additional uncertainty 
in the model.

\subsection{Nucleon and kaon resonances}
In selection of a set of baryon resonances that preferably describe 
the world's $p(\gamma, K^+)\Lambda$ data, one has to perform thousands 
of fits assuming all acceptable resonance combinations. To our knowledge, 
such a robust analysis has been performed by Adelseck and Saghai~\cite{AS},  
further extended by the Saclay-Lyon group~\cite{SL} and by the Gent 
group~\cite{RPR11} using more sophisticated technique in the data analysis 
based on a Bayesian inference method. Another data analysis in the multipole 
approach was performed by Mart and Sulaksono~\cite{MartSu} who considered 
resonances with the spin up to 9/2 with 93 free parameters performing 
the $\chi^2$ minimization fits to CLAS, SAPHIR, and LEPS data.
The Gent group made the Bayesian test of a huge number of nucleon resonance 
combinations and selected two sets of the resonances with 
highest evidence values. 
We have chosen one of these solutions, RPR-2011A~\cite{RPR11}, as the starting 
point in our analysis. The corresponding resonances are: N3, N4, N7, N9, P1, 
P2, P3, and P4 (see Tab.~\ref{TableAllRes} for the notation). 
Since we limit ourselves only to the $K^+\Lambda$ channel, there is no need 
to introduce $\Delta$ resonances which cannot decay to $K^+\Lambda$ due 
to isospin conservation. 

The four-star resonance N3 ($S_{11}(1535)$), which is of crucial importance 
for the description of $\eta$ photoproduction, lies below the 
K$\Lambda$ threshold, but its coupling to the K$\Lambda$ channel is possible 
due to its large width and predicted strong coupling to the strangeness 
sector. In the Bayesian analysis with the RPR model the N3 resonance was 
found to contribute with a moderate probability~\cite{RPR11} whereas 
in the isobar model its coupling strength to the K$\Lambda$ channel 
was found to be quite small~\cite{S1535}. 

In the KM and Gent isobar models, the N4, N6, and N7 established resonances 
were chosen along with the missing resonances P4 ($D_{13}(1875)$) and P1 
($P_{11}(1880)$), respectively. In the SL model~\cite{SL} for the K$\Lambda$ 
electroproduction, only the well established 
resonances N1, N7, and N8 were selected. The older RPR model, 
RPR2007~\cite{GentRPR07}, selected N4, N6, N7, P2, and P4 resonances. 
The resonances N1, N6, and N8 were ruled out in the new Bayesian analysis 
whereas N4 and N7 and the missing P1, P2, and P4 resonances were confirmed. 
Note that due to large decay widths of most resonances their contributions 
overlap each other resulting in interference among many states. 
This makes the analysis of the resonance content of the invariant amplitude 
difficult, and even though high-quality data are available it still brings 
uncertain results (several possible solutions).

In the past, nucleon resonances P3 and P5 were considered as a one state only. 
Recently, the Particle Data Group~\cite{PDG} decided to consider them as two separate 
states. Since both of these states have only two-star status, they are not 
included in the PDG Summary Tables.

In many studies the vector K$^*$ and pseudovector K$_1$ meson resonances were 
found to be important in the data description~\cite{SL,WJC} and  
are used in all realistic isobar models. We have therefore included 
them in the basic resonance set. Let us remind that these two states together 
with the kaon are the lowest poles in the K$^+$ and K$^*$ Regge trajectories 
included in the Regge~\cite{Guidal} and RPR~\cite{GentRPR07,RPR11} models, 
which also corroborates the importance of these states.

\subsection{Hyperon resonances}
The exchanges of hyperon resonances in the \emph{u}-channel contribute to 
the background and were not included in some isobar models, \emph{e.g.}, in KM.  
They can play, however, an important role in the dynamics as shown in the 
SL~\cite{SL} and Gent isobar~\cite{Jan01A,Jan01B} models. 
Particularly, they can compensate the nonphysically 
big contributions of the Born terms. Moreover, their presence can significantly 
improve description of data, reduce the $\chi^2$ and shift the value 
of the hadron cut-off parameter to a harder region~\cite{MartY}.

Formerly, mainly spin-1/2 hyperon resonances were included in the models 
with inconsistent description of the spin-3/2 baryons. 
To our knowledge, the only attempt to include a spin-3/2 hyperon 
resonance in the isobar model was done by the Saclay-Lyon group in Ref.~\cite{SLA},  
the version ``C'' of the SL model.
The reason 
for this limitation was that the pole in the $u=q^2$ variable, 
which appears in the invariant amplitude from the projection operator 
of the propagator (\ref{32propagator}), lies in the physical region ($u=0$)  
causing a divergence of the amplitude with the inconsistent interaction. 
In the consistent formalism, the pole term does not contribute   
owing to the transversality of the interaction vertexes, 
Eq.~(\ref{transversV}), and regularity of the amplitude,
\begin{equation}
V^{EM}_\mu(N^*p\,\gamma)\;\frac{\not\!q + m_R}{u-m_R^2+\texttt{i}m_R\Gamma_R}\, 
\frac{1}{3u}(\not\! q\,q^\nu \gamma^\mu + 
q^\mu \gamma^\nu\not\!q)\;V^S_\nu (K\Lambda N^*) = 0,
\label{eq:Y^*(3/2)}
\end{equation}
leaving only nonzero contributions from the momentum-independent 
terms in the projection operator (\ref{eq:32projector}). 
It is therefore safe to include the spin-3/2 hyperon resonances with 
relatively small masses, see Tab.I, which are expected 
to be important in describing the background.

Here we have considered only the spin-3/2 $\Lambda$ and $\Sigma$ well established 
four- or three-star resonances as reported in the Particle Data 
Tables 2014~\cite{PDG} (Table I): $\Lambda (1520)\,3/2^-$ (L6), 
$\Lambda (1690)\,3/2^-$ (L7), and $\Lambda (1890)\,3/2^+$ (L8), 
with the branching ratios to $N\bar{K}$ 45\%, 20-30\%, and 20-35\%, 
respectively; $\Sigma (1670)\,3/2^-$ (S3) and $\Sigma (1940)\,3/2^-$ (S4) 
with 7-13\% and $<$20\%, respectively.

\subsection{Hadron form factors}
\label{sect:hff}
Apart from reduction of the Born terms, the hadron form factor 
can also mimic the internal structure of hadrons in the strong vertexes, 
which is neglected in the hadrodynamical approach. However, there is still 
an ambiguity in the selection of a form of the hadron form 
factor: one can choose among dipole $F_d$, multidipole $F_{md}$, 
Gauss $F_{G}$, or multidipole Gauss shape $F_{mdG}$~\cite{Gent_spin}:

\begin{subequations}
\begin{align}
F_d (x,m_R,\Lambda_R) = & \,\frac{\Lambda_R^4}{(x-m_R^2)^2+\Lambda_R^4},\label{eq:Fd}\\
F_{md}(x,m_R,\Lambda_R,J_R) = & \,F_d^{J_R+1/2}(x,m_R,\Lambda_R),\label{eq:Fmd}\\
F_G (x,m_R,\Lambda_R) = & \,\exp [-(x-m_R^2)^2/\Lambda_R^4],\label{eq:FG}\\
\begin{split}F_{mdG}(x,m_R,\Lambda_R,J_R,\Gamma_R) = 
& \,F_d^{J_{R}-1/2}(x,m_R,m_R\tilde{\Gamma}_R) 
 \, F_G (x,m_R,\Lambda_R),\label{eq:FmG}\end{split}
\end{align}
\label{eq:FFall}
\end{subequations}
where $m_R$, $J_{R}$, $\Lambda_R$, and $x\equiv s,t,u$ stands for the mass 
and spin of the particular resonance, cut-off parameter of the form factor, 
and Mandelstam variables, respectively. Moreover, it is required to introduce 
a modified decay width 
\begin{equation}
\tilde{\Gamma}_R(J_R) = \frac{\Gamma_R}{\sqrt{2^{1/2J_R}-1},}
\end{equation}
which depends on the spin of the resonance and leads to preserving 
the interpretation of the resonance decay width as the full width in 
half maximum (FWHM) of the resonance peak~\cite{Gent_spin}.

\begin{figure}[h!]
\includegraphics[width=11cm]{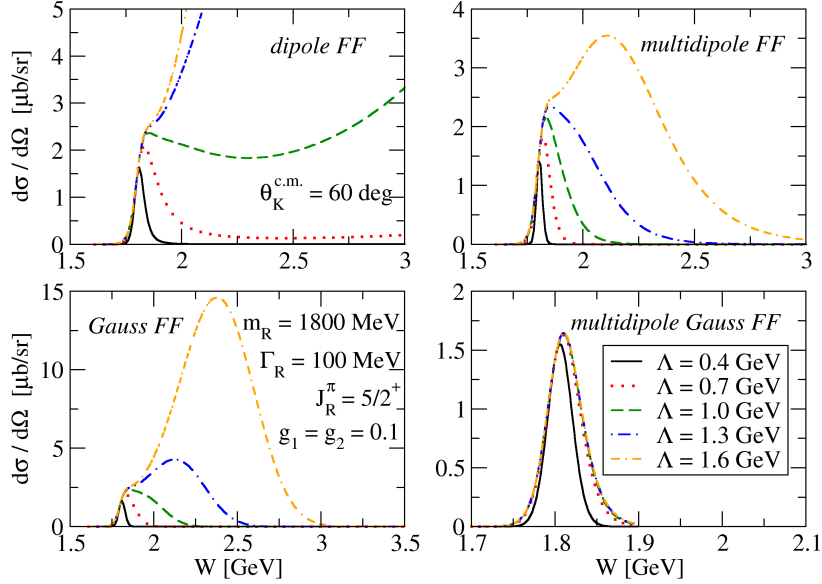}
\caption{(Color online) Contribution of the spin-5/2 resonance with the mass 1800 
and width 100~MeV to the cross section using different form factors. 
The cut-off dependence of the contribution is shown: 
the larger the cut-off value $\Lambda$, the more pronounced the effect.}
\label{fig:hff}
\end{figure}

Since the high-power momentum dependence of the amplitude leads to a substantial 
growth of the resonance contribution to the cross section, we need to 
introduce a hadron form factor to refine this behavior. In fact, the form factor 
should ensure that the resonant diagram does not contribute far from the mass pole 
of the exchanged particle. Unfortunately, with the form factor the cut-off 
dependence is introduced into the cross section. In Fig.~\ref{fig:hff}, we 
demonstrate the dependence for the contribution of a particular resonance 
with spin 5/2 in the \emph{s}-channel using the dipole (\ref{eq:Fd}), 
multidipole (\ref{eq:Fmd}), Gauss (\ref{eq:FG}), and 
multidipole Gauss (\ref{eq:Fmd}) form factors with various values of 
the cut-off parameters.
The use of the dipole form factor leads to enlarging the tail of 
the resonant peak whereas the Gauss form factor creates an 
artificial cut-off-value dependent peak while the actual resonant 
peak contributes only as its shoulder.
Introducing the spin-dependent form factor, multidipole or 
multidipole Gauss, makes the effect moderate even for larger values 
of the cut-off parameter. Using the latter form factor makes 
the contribution almost independent of the cut-off value producing 
the real resonance pattern in the cross section (see Fig. \ref{fig:hff}). 

The total amplitude constructed with the help of the effective 
Lagrangians is gauge invariant. The resonant amplitudes and 
the \emph{u}-channel Born contribution are gauge-invariant on 
their own, the gauge non-invariant terms occur in the \emph{s}- 
and \emph{t}-channel Born contributions, see Eqs. (\ref{eq:Bs-gauge}) 
and (\ref{eq:Bt-gauge}) in Appendix \ref{sect:contrib}. 
However, these terms cancel in the sum of these two Born 
contributions. Unfortunately, while introducing the hadron 
form factors, these gauge noninvariant terms no longer cancel. 
The remedy is to introduce a contact term which ensures the 
gauge invariance~\cite{Jan01A}, see Appendix \ref{sect:contact} 
for more details. 

The generally accepted cut-off values lie in the range from 
approximately $0.7$ to $3.0\,\mbox{GeV}$; 
the lower the cut-off, the stronger the suppression. 
The values around the lower limit are considered 
too soft, and the form factors are, in this situation, 
regarded as a rather artificial tool to suppress the Born term 
contribution. As our analysis showed, obtaining a harder cut-off 
value is much easier than a softer one, which we attribute to the 
presence of many hyperon resonances in background.

Values of the cut-off parameters are established 
when optimizing the model parameters against experimental data. 
A single common cut-off value $\Lambda_R$ is assumed for all 
resonant diagrams whereas for the background terms another value 
$\Lambda_{bgr}$ is used.

%
%
\section{Observables}
In the electroproduction 
$${\rm e}(k_1) + {\rm p}(p) \to {\rm e}(k_2) + {\rm K^+}(p_K) 
+ \Lambda (p_\Lambda),$$ 
the transition amplitude in the one-photon exchange approximation  
is a product of the matrix elements of the hadron $\mathbb{J}^\mu$ 
and lepton 
$l_\mu = e\,\bar{u}(k_2)\gamma_\mu u(k_1)$ currents mediated 
by the photon propagator 
\begin{equation}
{M}_{fi} = \frac{1}{k^2}\,l_\mu\,\mathbb{J}^\mu(k^2,s,t,u)\,, 
\end{equation}
where $k = k_1 - k_2$ is the four-momentum of the virtual photon 
and $s=(p+k)^2$, $t=(p_K-k)^2$, and $u=(p_\Lambda-k)^2$ 
are the Mandelstam variables. Conservation of the hadron  and lepton
currents implies 
$\mathbb{J}^\mu k_\mu = l^\mu k_\mu = 0$. 
The matrix element of the hadron current therefore can be decomposed into  
the linear combination of six covariant gauge-invariant contributions 
\begin{equation}
\mathbb{J}^\mu \varepsilon_\mu= \sum_{j=1}^{6}\,\mathcal{A}_{j}(k^2,s,t,u)\;
\bar{u}(p_{\Lambda})\,\gamma_{5}\,\mathcal{M}_{j}\,u(p)\,,
\label{eq:compact}
\end{equation}
where $\mathcal{M}_{j}$ are explicitly gauge-invariant operators 
\begin{subequations}
\begin{align}
\mathcal{M}_{1} &=(\not \!k \not\!\varepsilon - \not\!\varepsilon \not \!k )/2,\\
\mathcal{M}_{2} &= p\cdot\varepsilon-k\cdot p\, k\cdot\varepsilon/k^{2},\\
\mathcal{M}_{3} &= p_{\Lambda}\cdot\varepsilon - k\cdot p_{\Lambda}\, k\cdot\varepsilon/k^{2},\\
\mathcal{M}_{4} &= \not\!\varepsilon k\cdot p - \not \! k p\cdot\varepsilon,\\
\mathcal{M}_{5} &= \not\!\varepsilon k\cdot p_{\Lambda} - \not \!k p_{\Lambda}\cdot\varepsilon,\\
\mathcal{M}_{6} &= \not \! k k\cdot\varepsilon - \not\!\varepsilon k^{2},
\end{align}
\end{subequations}
and $\varepsilon_\mu$ is the polarization vector of the virtual photon.
The scalar amplitudes $\mathcal{A}_{j}(k^2,s,t,u)$ contain contributions 
from the considered tree-level Feynman diagrams. Their expressions for various types of 
particle exchanges are given in Appendix \ref{sect:contrib}. 
In the photoproduction case ($k^2=0$), there are only four terms  
in the decomposition (\ref{eq:compact})~\cite{AS}.

In the calculations which involve also a non relativistic input, 
\emph{e.g.},  the calculation of the hypernucleus production cross 
sections~\cite{HProd} with non relativistic wave functions of 
the nucleus and hypernucleus, one also needs a more convenient 
representation of the Lorentz invariant matrix element (\ref{eq:compact}) 
in terms of the two-component spinor 
amplitudes known as the Chew, Goldberger, Low, and Nambu (CGLN) 
amplitudes~\cite{AS,SL,SF}. These amplitudes are, however, also widely 
used in calculations of observables in the elementary process.
In the c.m. frame, the Lorentz invariant matrix element (\ref{eq:compact}) can be written as 
\begin{equation}
\mathbb{J}^\mu\varepsilon_\mu = 
\chi_\Lambda^+\, {\cal F}\, \chi_{\rm p}
\end{equation}
where $\chi_{\rm p}$ and $\chi_\Lambda$ are the Pauli spinors and 
\begin{equation}
{\cal F} =  f_1\,\vec{\sigma}\cdot\vec{\varepsilon} 
- \texttt{i}f_2\,\vec{\sigma}\cdot \hat{\vec{p}}_K \, 
  \vec{\sigma}\cdot(\hat{\vec{k}}\times\vec{\varepsilon})
+ f_3\,\vec{\sigma}\cdot\hat{\vec{k}}\,\hat{\vec{p}}_K\cdot \vec{\varepsilon} 
+ f_4\,\vec{\sigma}\cdot\hat{\vec{p}}_K\,\hat{\vec{p}}_K\cdot\vec{\varepsilon}
+ f_5\,\vec{\sigma}\cdot\hat{\vec{k}}\,\hat{\vec{k}}\cdot\vec{\varepsilon} 
+ f_6\,\vec{\sigma}\cdot\hat{\vec{p}}_K\,\hat{\vec{k}}\cdot\vec{\varepsilon}.
\end{equation}
Here $\hat{\vec{k}}=\vec{k}/|\vec{k}|$, $\hat{\vec{p}}_K=\vec{p}_K/|\vec{p}_K|$, 
$\vec{\sigma}$ are the Pauli matrices, and $\vec{\varepsilon}$ is the spatial 
component of the virtual-photon polarization vector. 
The CGLN amplitudes $f_i(k^2,s,t,u)$ are expressed via the scalar 
amplitudes $A_j$
\begin{subequations}
\begin{align}
f_{1}&=N^* [-(W-m_p)\mathcal{A}_{1}+k\cdot p\,\mathcal{A}_{4} 
+ k\cdot p_{\Lambda}\,\mathcal{A}_{5}-k^{2}\mathcal{A}_{6}],\\
f_{2}&=N^* \frac{|\vec{k}||\vec{p}_{K}|}{(E^*_{\Lambda}+m_{\Lambda})(E_p^* 
+ m_{p})}[(W+m_p)\mathcal{A}_{1} 
+ k\cdot p\,\mathcal{A}_{4}+k\cdot p_{\Lambda}\,\mathcal{A}_{5}-k^{2}\mathcal{A}_{6}],\\
f_{3}&=-N^* \frac{|\vec{k}||\vec{p}_{K}|}{E_{p}^*+m_{p}}[\mathcal{A}_{3}+(W + 
m_{p})\mathcal{A}_{5}],\\
f_{4}&=N^*\frac{|\vec{p}_{K}|^{2}}{E_{\Lambda}^*+m_{\Lambda}}[\mathcal{A}_{3}-(W-m_{p})\mathcal{A}_{5}],\\
f_{5}&=N^*\frac{|\vec{k}|^{2}}{E_{p}^*+m_{p}}\bigg[\mathcal{A}_{1}-\frac{1}{k^{2}}
[(k^{2}+k\cdot p)\mathcal{A}_{2}
+ k\cdot p_{\Lambda}\,\mathcal{A}_{3}]-(W+m_{p})(\mathcal{A}_{4}+\mathcal{A}_{6})\bigg],\\
\begin{split}f_{6}&=N^*\frac{E_{\gamma}^*|\vec{k}||p_{K}|}{(E^*_{\Lambda}+m_{\Lambda})(E^*_{p}+m_{p})}\bigg\{\mathcal{A}_{1}-m_{p}\mathcal{A}_{4} 
+ \frac{k\cdot p_{\Lambda}}{E_{\gamma}^*}\mathcal{A}_{5}+\frac{(E^*_{p}+m_{p})}{E_{\gamma}^* k^{2}}[(k^{2}+k\cdot p)\mathcal{A}_{2}\\
&\;\;\;\;+k\cdot p_{\Lambda}\,\mathcal{A}_{3}]-(W+m_p)\mathcal{A}_{6}\bigg\},\end{split}
\end{align}
\end{subequations}
where $W=\sqrt{s}$ and $E_p^*$, $E_\Lambda^*$, $E_{K}^*$, and $E_{\gamma}^*$ are 
the c.m. energies of the proton, hyperon, kaon, and photon, respectively. 
The normalization factor reads
\begin{equation}
N^* = \sqrt{\frac{(E^*_{\Lambda}+m_{\Lambda})(E^*_{p}+m_{p})}{4m_{\Lambda}m_{p}}}.
\end{equation}

The triple-differential cross section for electroproduction of unpolarized 
hyperon with unpolarized electron beam and target is obtained as
\begin{equation}
\frac{\mbox{d}^{3}\sigma}{\mbox{d}E_{e'}\mbox{d}\Omega_{e'}\mbox{d}\Omega_{K}^{c.m.}}
= \Gamma\bigg[\sigma_{T}+\varepsilon \sigma_{L}+\varepsilon \sigma_{TT}\cos 2\phi_{K} 
+ \sqrt{2\varepsilon_{L}(\varepsilon+1)}\sigma_{LT}\cos\phi_{K}\bigg],
\label{eq:elcrs}
\end{equation}
where $\phi_{K}$, $\Gamma$, $\varepsilon$, and $\varepsilon_{L}$ are  
the angle between the lepton and hadron planes, the virtual-photon 
flux factor, and the transverse and longitudinal photon polarization 
parameters, respectively~\cite{SF}.  
The response functions $\sigma_T$ and $\sigma_L$ describe the cross 
sections for the unpolarized and longitudinally polarized photon beams, 
respectively, while $\sigma_{TT}$ stands for the asymmetry of 
a transversally polarized photon beam. The last term containing $\sigma_{LT}$ 
describes the interference effects between the longitudinal and transverse 
components of the photon beam. Note that $\sigma_{T}$ and $\sigma_{TT}$ correspond 
to the cross section and beam asymmetry in the photoproduction process, respectively.
The response functions in terms of the CGLN amplitudes read as follows
\begin{subequations}
\begin{align}
\sigma_{T}=&\,C\, \texttt{Re}\bigg\{|f_{1}|^{2}+|f_{2}|^{2} - 
2f_{1}f_{2}^{*}\cos \theta_{K}
+ \sin^{2}\theta_{K}\bigg[\frac{1}{2}(|f_{3}|^{2}+|f_{4}|^{2})+f_{1}f_{4}^{*}
+ f_{2}f_{3}^{*}+f_{3}f_{4}^{*}\cos\theta_{K}\bigg]\bigg\},\\
\sigma_{L}=&\,C\, \texttt{Re}\left\{|\tilde{f}_{5}|^{2}+|\tilde{f}_{6}|^{2}+2\tilde{f}_{5}\tilde{f}_{6}^{*}\cos\theta_{K}\right\},\\
\sigma_{TT}=&\,C\, \texttt{Re}\bigg\{\frac{1}{2}(|f_{3}|^{2}+|f_{4}|^{2})+ 
f_{1}f_{4}^{*}+f_{2}f_{3}^{*}
+ f_{4}f_{3}^{*}\cos\theta_{K}\bigg\}\sin^{2}\theta_{K},\label{sigmaTT}\\
\sigma_{LT}=&-C\, \texttt{Re}\bigg\{(f_{1}+f_{4})\tilde{f}_{6}^{*}+ 
(f_{2}+f_{3})\tilde{f}_{5}^{*} 
+ (f_{3}\tilde{f}_{6}^{*}+f_{4}\tilde{f}_{5}^{*})\cos\theta_{K}\bigg\}\sin\theta_{K},
\end{align}
\end{subequations}
where we have defined the linear combinations
\begin{align}
\tilde{f}_5 =& f_1 + f_3 \cos\theta_{K} + f_5,\\
\tilde{f}_6 =& f_4 \cos \theta_K + f_6
\end{align}
and the normalization factor $C$ is given as
\begin{equation}
C = (\hbar c)^2 \frac{\alpha}{4\pi} \frac{m_{\Lambda}|\vec{p}_{K}|}{|\vec{k}|W}.
\end{equation}
The general expression for the electroproduction cross section considering 
all three possible types of polarization can be found in Ref.~\cite{Koch}. 
Here we give only the single-polarization observables in photoproduction 
which we use in the analysis and which in terms of the CGLN amplitudes read
\begin{align}
P &= -\texttt{Im}[2f_1^*f_2+f_1^*f_3-f_2^*f_4-(f_2^*f_3-f_1^*f_4)\cos\theta_K 
- f_3^*f_4\sin^2\theta_K]\sin \theta_K, \\
\Sigma &= -\texttt{Re}[(|f_3|^2+|f_4|^2)/2+f_2^*f_3+f_1^*f_4
+ f_3^*f_4\cos\theta_K]\sin^2\theta_K,\\
T &=\texttt{Im}[f_1^*f_3-f_2^*f_4+\cos\theta_K(f_1^*f_4-f_2^*f_3)
- f_3^*f_4\sin^2\theta_K]\sin\theta_K,
\end{align}
where $P$, $\Sigma$ and $T$ stands for hyperon polarization, beam asymmetry (see 
also Eq.~(\ref{sigmaTT})) and target polarization, respectively.
%
%
%
\section{Fitting model parameters} 
Since the isobar model is an effective model with the coupling constants and cut-off 
values of hadron form factors undetermined, our goal is to fixate these free 
parameters to the experimental data during the fitting process. 

The free parameters to be adjusted are the coupling constants of the Born terms 
$g_{K\Lambda N}$ and $g_{K\Sigma N}$, the nucleon, kaon and hyperon resonances 
and two cut-off parameters of the hadron form factor. Each spin-1/2 resonance 
contributes with one parameter whereas higher-spin resonances as well as kaon 
resonances contribute with two parameters. As well as in the well-known Kaon-MAID 
model, we assume a single cut-off value $\Lambda_R$ for all 
resonant (\emph{s}-channel) diagrams whereas for background terms another 
value $\Lambda_{bgr}$ is used. Altogether, the number of free parameters 
varies from 20 to 25 depending on the number and spin of considered nucleon 
and hyperon resonances.

In order to test whether a given hypothetical function describes the given 
data well, the $\chi^2$ is calculated. The optimum set of free parameters 
$(c_1,\ldots, c_n)$ for a given set of data points $(d_1,\ldots, d_N)$ is 
that with the lowest value of $\chi^2$. The $\chi^2$ is
\begin{equation}\chi^2 = \sum_{i=1}^{N}\frac{[d_{i}-p_{i}(c_{1},\ldots,c_n)]^2}{(\sigma_{d_{i}}^{tot})^{2}},
\end{equation}
where $N$ is the number of data points and $n$ the number of free parameters; 
$p_i$ represents the theoretical prediction of observables (differential 
cross section, hyperon polarization and beam asymmetry in our case) for 
the measured data point $d_i$, with the total error given as
\begin{equation}
\sigma_{d_{i}}^{tot}=\sqrt{(\sigma_{d_{i}}^{sys})^{2}+(\sigma_{d_{i}}^{stat})^{2}},
\label{eq:sigma-tot}
\end{equation}
where $\sigma_{d_{i}}^{sys}$ and $\sigma_{d_{i}}^{stat}$ represent systematic 
and statistical errors of a given datum, respectively. Whereas systematic 
errors tend to be strongly correlated within a given data set, the correlation 
weakens when using several independent subsets. Since we assume several data 
sets (see subsection \ref{sect:data}), we have adopted the 
definition (\ref{eq:sigma-tot}) similarly to the analysis by Adelseck and 
Saghai \cite{AS}. Some groups, \emph{e.g.}, the Gent group, use an even more 
conservative prescription for the total error~\cite{RPR11}. 

In order to obtain the optimum set of parameters, one is forced to 
minimize $\chi^2$ in the $n$ dimensional space. In the ideal case,
$\chi^2 = \text{n.d.f.}$,
where $\text{n.d.f.} = N - n$ is the number of degrees of freedom.

The minimization was performed with the help of the least-squares fitting procedure 
using the MINUIT code~\cite{CERN}. Since MINUIT uses a nonlinear transformation 
for the parameters with limits, which makes the accuracy of the resulting parameter worse when it approaches a boundary value, the limits should be avoided if they are not 
necessary to prevent the parameter from reaching nonphysical values. The main 
coupling constants $g_{K\Lambda N}$ and $g_{K\Sigma N}$ were kept inside the 
limits of 20\% broken SU(3) symmetry
\begin{subequations}
\begin{align}
-4.4\leq &\,\frac{g_{K\Lambda N}}{\sqrt{4\pi}}\leq -3.0,\label{eq:SU3a}\\
0.8 \leq &\, \frac{g_{K\Sigma N}}{\sqrt{4\pi}}\leq 1.3.
\end{align}
\label{eq:SU3}
\end{subequations}

In order to avoid too soft or too hard form factors, the cut-off parameters of 
the hadron form factor were kept inside the limits from $0.7$ to 
$3.0\,\mbox{GeV}$.

The coupling parameters entering the fitting procedure are always products of the 
strong and electromagnetic coupling constants. In order to guarantee a correct 
dimension of the interacting Lagrangians, the coupling constants have to be 
normalized appropriately. Since the Lagrangian for the spin-3/2 nucleon resonance  
contains two derivatives of the R-S field, the coupling parameters read
\begin{subequations}
\begin{align}
G_{1}=& \frac{f g_{1}}{m_{R}^{2}m_{K}(m_{R}+m_{p})},  \\
G_{2}=&\frac{f g_{2}}{m_{R}^{2}m_{K}(m_{R}+m_{p})}.
\end{align}
\label{eq:ccN32}
\end{subequations}
In the case of spin-3/2 hyperon resonances, $m_p$ is replaced with $m_\Lambda$. 
Analogously, the spin-5/2 coupling parameters are normalized as follows
\begin{subequations}
\begin{align}
G_1 =& -\frac{f g_{1}}{16 m_K^4 m_p^4},\\ 
G_2 =& -\frac{f g_{2}}{32m_K^4 m_p^5}.
\end{align}
\label{eq:ccN52}
\end{subequations}
The high mass powers in the denominator  result in very small values of $G_i$ for 
$N^*(5/2)$ in comparison with the coupling parameters of lower-spin nucleon resonances.

The hyperon coupling parameters tend to be very large compared with coupling 
parameters of other resonances. Therefore, we did not take into account results 
with hyperon coupling parameters significantly bigger than 10. 
         
\subsection{Experimental data}
\label{sect:data}
Recently, new precise data from LEPS, GRAAL and particularly from CLAS 
collaborations became available. For the fitting procedure, we selected 
around 3400 data points stemming from CLAS and LEPS collaborations with 
addition of several tens of data points collected by Adelseck and 
Saghai~\cite{AS}. Namely, we used the CLAS 2005~\cite{CLAS05}, 
CLAS 2010~\cite{CLAS10}, and LEPS~\cite{LEPS} cross-section data, 
CLAS 2010 hyperon polarization data~\cite{CLAS10}, and LEPS beam 
asymmetry data~\cite{LEPS}. 

In our analysis, we are concerned mainly with the resonance region and 
therefore have restricted the CLAS 2010 data sets to the energy range up 
to $2.355\,\mbox{GeV}$ and $2.225\,\mbox{GeV}$ for the cross-section and hyperon
polarization data, respectively.

Since the CLAS and SAPHIR~\cite{SAPHIR} data are not consistent with each 
other, especially in the forward-angle region~\cite{ByMa} which is of particular 
interest here, we decided not to use the SAPHIR cross-section data in the analysis. 
Unfortunately, the CLAS 2005 and CLAS 2010 data sets show inconsistency with each 
other of about one or two standard deviations in the threshold region for kaon 
angle less than approximately $60^\circ$.

\subsection{Results of fitting}
While minimizing the $\chi^2$ it is important to find a global minimum. Since this 
task occurs in a huge parameter space that has a lot of local minima, the result 
of the fitting procedure often depends on starting values of the fitted parameters.

Generally, choosing the best solution is not an easy task. The $\chi^2$ value is 
only a mathematical tool showing the goodness of a fit. However, results with 
similar $\chi^2$ values can still give rather different predictions of the observables 
in some kinematic regions. Therefore, not only thorough inspection of the numerical 
values of the fitted parameters, but also a brief check of the predicted observables 
is welcome.

We have done several hundreds of fits considering various resonance configurations 
and different shapes of hadron form factor. While the set of nucleon resonances 
chosen in the RPR-2011A model provided us the starting point, we have considered 
many other resonant states during the procedure of fitting.

Since one cannot be sure that the detected minimum is the global one, we have 
selected several models with similar $\chi^2$. The models differ mainly in the 
choice of nucleon and hyperon resonances and their coupling constants, cut-off 
values of the hadron form factor, and the shape of the form factor. Particularly, 
the smallness of hyperon coupling constants plays an important role when deciding 
if the model should be rejected or not. Since the isobar model is only a tree-level 
approximation, the couplings even larger than 1 are still justifiable. 

During the fitting procedure, we also tried to slightly modify the mass and 
width of several intermediate particles in the ranges provided by the Particle Data 
Tables 2014~\cite{PDG} (or when there were no preferred values). On the one hand, 
this forced the models to improve their description of the cross section; 
especially the reduction of the width of P2 from $500\,\mbox{MeV}$ to a value 
of about $400\,\mbox{MeV}$ or less led to filling up the second peak in the 
cross-section data. On the other hand, the modification of the width of P2 
resulted in a growth of the $\chi^2$ value and made the description of 
single-polarization observables worse. 

In order to gain insight into the effect of high-spin resonances on the observables, 
some of the fits were performed with the inconsistent formalism for the spin-3/2 
and spin-5/2 resonances used in the SL model. Particularly, the fit of the BS2 model 
(see below) with the inconsistent formalism led to an enlargement of the $\chi^2$ 
from $1.64$ to $1.91$, growth of the cut-off parameter for the hadron form factor to 
almost $3\,\mbox{GeV}$, and decrease of the cross-section prediction in the 
forward-angle region. In this fit we omitted the spin-3/2 hyperon resonance S4. 
Generally, the use of the inconsistent high-spin formalism results in larger 
couplings for spin-5/2 resonances which is due to a different normalization 
introduced into the coupling parameters (see Eq.~(\ref{eq:ccN52})).

The main asset of the presence of high-spin hyperon resonances is the reduction 
of coupling parameters of spin-1/2 hyperon resonances. With no $Y^*(3/2)$ 
introduced, the couplings of $Y^*(1/2)$ tend to acquire values of the order 
of 10 or even more. While the $Y^*(3/2)$ are implemented, the couplings of 
both $Y^*(1/2)$ and $Y^*(3/2)$ are only exceptionally bigger than 10.

In the analysis, we examined the effect of distinct shapes of the hadron form factor on the resonance behavior. As seen from the definition (\ref{eq:FFall}), the multidipole form factor affects the resonance behavior more strongly than the dipole one. Therefore, introducing the multidipole form factor leads to bigger cut-off parameters for resonances ($\Lambda_{res} \sim 3\,\mbox{GeV}$) than considering the dipole form factor ($\Lambda_{res} \sim 2\,\mbox{GeV}$). Unfortunately, we were not able to achieve a single result with $\chi^2<2$ using the multidipole Gauss form factor. This shape of form factor was introduced by the Gent group in their Regge-plus-resonance model to strongly suppress the contribution of the nucleon resonances in the high-energy region. However, it seems there is no need to introduce such a strong form factor in the isobar model.

The predictions of the models with $\chi^2<2$ were tested in the comparison with the experimental data. Particularly, the comparison with hyperon polarization data can reveal a subtle interplay among many resonances. Even though the smallness of $\chi^2$ denotes a good agreement of the model prediction with the data, in the kinematic regions where data are scarce (\emph{e.g.}, the forward-angle region) the model predictions can still differ. 

The best solutions regarding the smallness of the $\chi^2$, values of fitted parameters and correspondence with data were coined BS1 and BS2. Whereas the model coined BS1 was obtained using a multidipole form factor, the BS2 model was gained using a dipole shape of the form factor. Moreover, the mass of the P5 resonance was slightly modified from $1820\,\mbox{MeV}$ in the BS1 model to $1860\,\mbox{MeV}$ in the BS2 model.

\begin{table}[h!]
\begin{center}
\begin{tabular}{r||r r r r}
		                &~BS1~      &~BS2~      &~KM~     &~SL~   \\ \hline
 $g_{K\Lambda N}$       &$-3.00$    &$-3.00$    & $-3.80$ &$-3.16$\\
 $g_{K\Sigma^0 N}$      &$ 1.11$    &$ 0.80$    & $ 1.20$ &$ 0.91$\\
 $G_V(K^*)$             &$-0.18$    &$-0.17$    & $-0.79$ &$-0.05$\\
 $G_T(K^*)$             &$ 0.02$    &$-0.03$    & $-2.63$ &$ 0.16$\\
 $G_V(K_1)$              &$ 0.28$    &$ 0.30$    & $ 3.81$ &$-0.19$\\
 $G_T(K_1)$              &$-0.28$    &$-0.23$    & $-2.41$ &$-0.35$\\
 $G(N1)$                &   --      &   --      & --      &$-0.02$\\
 $G(N3)$                &$ 0.10$    &$ 0.17$    & --      & --    \\
 $G(N4)$                &$-0.07$    &$-0.05$    & $-0.13$ & --    \\
 $G(N6)$                &   --      &$-0.05$    & $-0.26$ & --    \\
 $G_1(N7)$              &$-0.09$    &$-0.07$    & $ 0.05$ &$-0.04$\\
 $G_2(N7)$              &$-0.01$    &$-0.0057$  & $ 0.61$ &$-0.14$\\ 
 $G_1(N8)$              &   --      &   --      &    --   &$-0.63$\\
 $G_2(N8)$              &   --      &   --      &    --   &$-0.05$\\ 
 $G_1(P4)$              &$ 0.21$    &$ 0.23$    & $ 1.10$ & --    \\
 $G_2(P4)$              &$ 0.26$    &$ 0.26$    & $ 0.63$ & --    \\
 $G_1(P5)$              &$-0.04$    &$-0.02$    & --      & --    \\
 $G_2(P5)$              &$ 0.04$    &$ 0.02$    & --      & --    \\
 $G_1(P2)$              &$ 0.11$    &$ 0.09$    & --      & --    \\
 $G_2(P2)$              &$-0.02$    &$-0.01$    & --      & --    \\
 $G_1(P3)$              &$-0.0003$  &$-0.0018$  & --      & --    \\
 $G_2(P3)$              &$-0.0029$  &$-0.0015$  & --      & --    \\
 $G_1(N9)$              &$ 0.05$    &$ 0.03$    & --      &$-0.63$\\
 $G_2(N9)$              &$-0.05$    &$-0.03$    & --      &$-0.05$\\
 $G(L1)$                & --        &$ 9.67$    & --      &$-0.42$\\
 $G(L3)$                & --        & --        & --      &$ 1.75$\\
 $G(L4)$                &$-8.39$    &$-11.55$   & --      & --    \\
 $G(L5)$                & --        & --        & --      &$-1.96$\\
 $G_1(L6)$              &$ 0.86$    & --        & --      & --    \\
 $G_2(L6)$              &$-0.09$    & --        & --      & --    \\ 
 $G_1(L8)$              &$-2.33$    & --        & --      & --    \\
 $G_2(L8)$              &$ 0.0033$  & --        & --      & --    \\
 $G(S1)$                &$-11.58$   &$-8.09$    & --      &$-7.33$\\
 $G(S2)$                &$ 15.77$   & --        & --      & --    \\
 $G_1(S4)$              &$-8.32$    &$-0.86$    & --      & --    \\
 $G_2(S4)$              &$ 0.81$    &$ 0.18$    & --      & --    \\ 
 $\Lambda_{bgr}$        &$ 1.88$    &$ 1.94$    & $ 0.64$ & --    \\
 $\Lambda_{res}$        &$ 2.74$    &$ 2.15$    & $ 1.04$ & --    \\ \hline
 $\chi^2/\text{n.d.f.}$ &$ 1.64$    &$ 1.64$    & --      & --    \\

\end{tabular}
\end{center}
\caption{Coupling constants, cut-off values, and $\chi^2$ 
of the final models BS1 and BS2 are compared with the parameters 
of the Kaon-MAID and Saclay-Lyon models.}
\label{tab:Y(3/2)}
\end{table}

%
%
\section{Discussion of results}  
In this section we present the new isobar models BS1 and BS2 for photoproduction 
of K$^+\Lambda$ and compare their predictions for the cross section, hyperon polarization, 
and beam asymmetry with the data and results of the older models Saclay-Lyon and 
Kaon-MAID. Note that the numerical results of the SL and KM models have been obtained 
by using our code with the parameters presented in Table~\ref{tab:Y(3/2)}.

The nucleon-resonance content of the BS1 and BS2 models almost does not differ, see 
Table~\ref{tab:Y(3/2)}. The BS2 contains only one more resonance N6 with a small 
coupling constant. The coupling constants of the other nucleon resonances have  
the same sign and their values are very similar. This set of $N^*$ significantly 
overlaps with that suggested by the Gent group. The only difference, except for N6, 
is that the two-star resonance P1 with spin 1/2 in the RPR was replaced with 
the almost equal mass two-star spin-5/2 resonance P5 in our models.

More differences between BS1 and BS2 are observed in description 
of the background. The values of the main coupling constants, 
$g_{K\Lambda N}$ and $g_{K\Sigma N}$, and those for $K^*$ 
and $K_1$ exchanges are very similar and the signs are 
identical except for the tensor coupling of $K^*$ which has 
the opposite sign. In both models the value of $g_{K\Lambda N}$ 
is at the upper limit allowed in fitting (\ref{eq:SU3a}), 
which suggests a considerable violation of SU(3) symmetry. 
Note that the differences in these coupling constants, particularly 
$g_{K\Sigma N}$ and $G_T(K^*)$, might have an impact on the model 
predictions in the $n(\gamma, K^0)\Lambda$ process~\cite{HYP10}.

Significant differences are found in the included sets of the hyperon resonances 
and their couplings. The BS2 contains only one spin-3/2 hyperon resonance S4 and 
three spin-1/2 resonances L1, L4, and S1, whereas BS1 includes three spin-3/2 
resonances L6, L8, and S4 and only one spin-1/2 resonance L4.  
The general feature of the presented models and other solutions 
found during the fitting procedure is that the coupling 
strengths of the hyperon exchanges tend to be relatively 
large in comparison with the typical values obtained for 
the couplings of the nucleon resonances. This experience is 
similar to that gained in the analyses by the Saclay-Lyon~\cite{SL} and 
Gent~\cite{Jan01B} groups on a role of the hyperon resonances 
in $p(\gamma, K^+)\Lambda$. 
Note that in version C of the Saclay-Lyon model~\cite{SLA} 
the only $\Lambda(1890)\,$3/2$^+$ (L8) resonance was included; they concluded, however,  
that this resonance is not required by the data set available at that time, 
i.e. before 1998.  
Reasonable values of the hyperon couplings, $-20 \le G(Y^*) \le 20$, 
were therefore used in our analysis as a criterion for a model selection.
These observations suggest that, whereas the current new 
experimental data are able to fix relatively well the set 
of the nucleon resonances producing genuine resonance 
patterns in the observables, they still cannot determine 
uniquely the non-resonant part of the amplitude (background).
Therefore, one still cannot select a set of hyperon resonances,  
contributing to the process, with certainty.

Let us note that, in view of the achieved quality of data 
description, the total number of resonances included in 
BS1 and BS2, 16 and 15, respectively, is quite moderate 
in comparison with the older models KM and SL and the 
recent models by Mart~\cite{MartSu,Mart} and Maxwell~\cite{Maxwell}.

\begin{figure}[t]
\includegraphics[width=12cm]{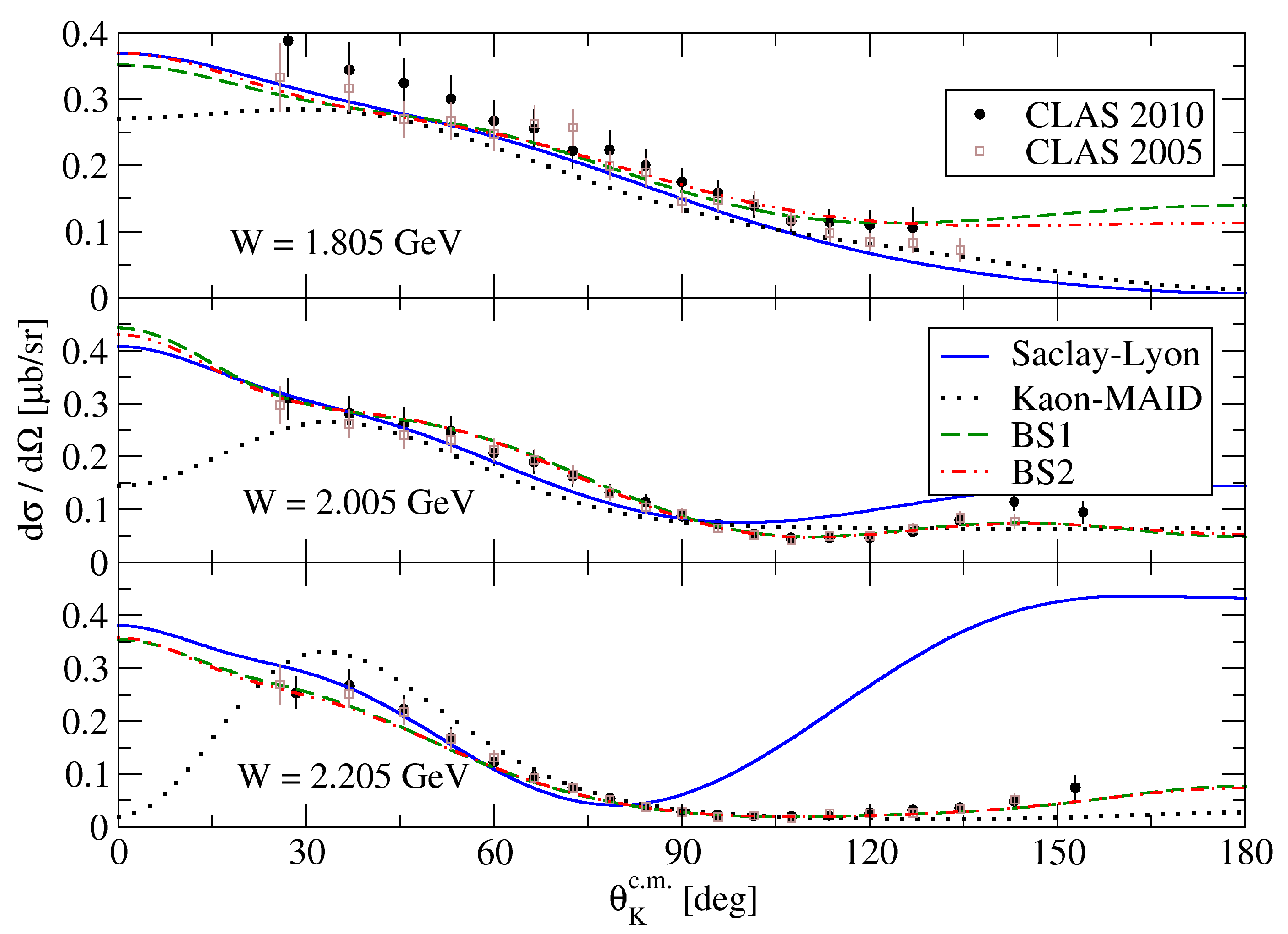}
\caption{(Color online) Angular dependence of the cross sections is shown 
for three values of the c.m. energy. In the forward-angle 
region, the Saclay-Lyon (solid curve), BS1 (dashed curve), 
and BS2 (dash-dotted curve) models predict decreasing dependence 
of the cross section. In contrast the Kaon-MAID model 
(dotted curve) predicts a bump around $\theta_K^{c.m.}=30^\circ$. 
The data are from the CLAS 2005~\cite{CLAS05} and CLAS 2010~\cite{CLAS10} 
collaborations.}
\label{fig:crs-theta}
\end{figure}
Angular dependence of the calculated cross sections in 
comparison with the CLAS data is shown in 
Fig.~\ref{fig:crs-theta} for three energies. Both BS1 and 
BS2 models give very similar predictions which differ from 
predictions of the other models mainly in the forward- 
and backward-angle regions. In the small kaon-angle 
region, $\theta_K^{c.m.}<40^\circ$, the new models predict  
descending angular dependence like the SL model, contrary to 
the KM which predicts a very suppressed cross sections for 
energies $W\ge 2$~GeV. In the backward-angle region the models 
agree with the KM describing the data very well. 
The subtle difference between BS1 and BS2 model in the description 
of backward angles (apparent for $W=1.805\,\mbox{GeV}$) can be 
assigned to the sign change of the tensor coupling of $K^*$.
One may conclude that the BS1 and BS2 models describe 
the cross sections in the full angular and considered 
energy regions very well. Note that the consistency of 
the cross sections in the very  small kaon-angle region 
with the results of the SL model (Fig.~\ref{fig:crs-theta}) 
and the fact that these cross sections are dominated by 
the spin-flip part of the amplitude could predetermine the new models 
for successful predictions of the cross sections in 
the production of the hypernuclei, like the Saclay-Lyon 
model~\cite{HProd,JLab}.

The model dynamics in the small-angle area is driven mainly  
by the background contributions in which the spin-1/2 hyperon 
resonances, surprisingly, play a very important role. 
In spite of their large contribution in the backward angles, 
they give the largest contribution in the forward-angle 
region when combined with the Born terms. 
On the other hand, the spin-3/2 hyperon resonances combined 
with the Born terms contribute predominantly in the 
backward-angle region. The role of the kaon resonances is 
to suppress the Born term contributions in the central-angle 
region.
\begin{figure}[t]
\includegraphics[width=12cm]{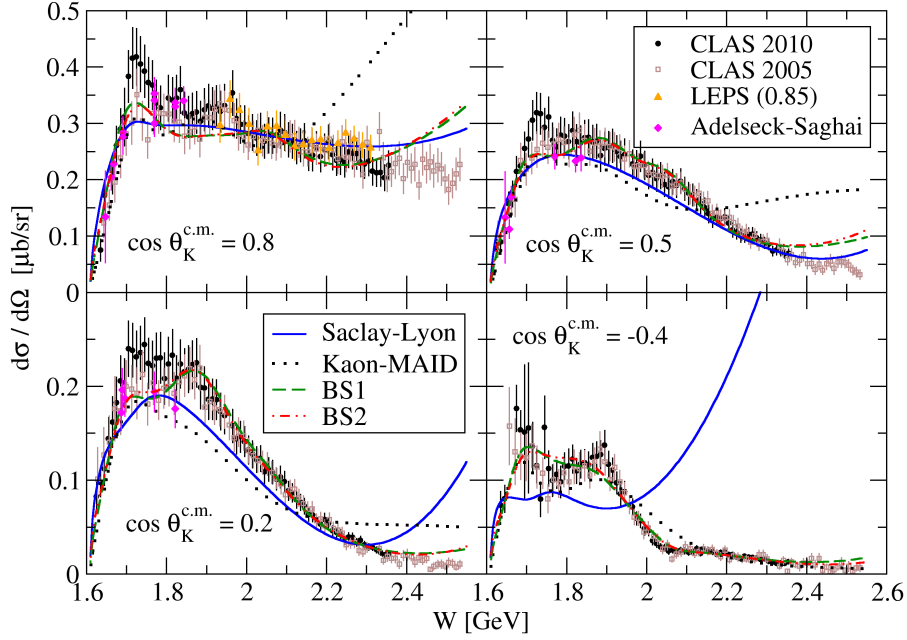}
\caption{(Color online) Differential cross section in dependence on 
the c.m. energy $W$ is shown for various kaon angles. 
Notation of the curves is the same as in the Fig. \ref{fig:crs-theta}. 
The data are from CLAS 2005~\cite{CLAS05}, CLAS 2010~\cite{CLAS10}, 
LEPS~\cite{LEPS} and from the publication of Adelseck and 
Saghai~\cite{AS}. The LEPS data are for $\cos \theta_K^{c.m.}=0.85$.}
\label{fig:crs-w}
\end{figure}

In Fig.~\ref{fig:crs-w} we show resonance effects in the 
energy dependent differential cross sections for four kaon angles as 
they are revealed by the data and the models. First, let 
us note that the resonance pattern revealed by the CLAS 
data around $W=1.7$~GeV for the forward angles is sharper 
in the CLAS data set from 2010 than in the older one from 2005. 
The new models predict conservative cross sections 
lying in between these data sets preferring rather the older 
data. The N6 in BS2 is not strong enough to make the peak around 
1.7~GeV sharper. The older CLAS data set is also favored by the 
hybrid RPR-2011A and RPR-2011B models~\cite{RPR11}. 
Both new isobar models BS1 and BS2 predict a peak around 
1.9~GeV in the central- and backward-angle regions but not 
at very small angles. In the forward-angle region some 
strength is also apparent around $W=2$~GeV modeled by the 
higher-mass resonances P2, P3, and P4. 
The strong grow of the cross section in the threshold region 
is described by the BS1 and BS2 models satisfactorily, better 
than by the SL model.

\begin{figure}[t]
\includegraphics[width=8cm,angle=270]{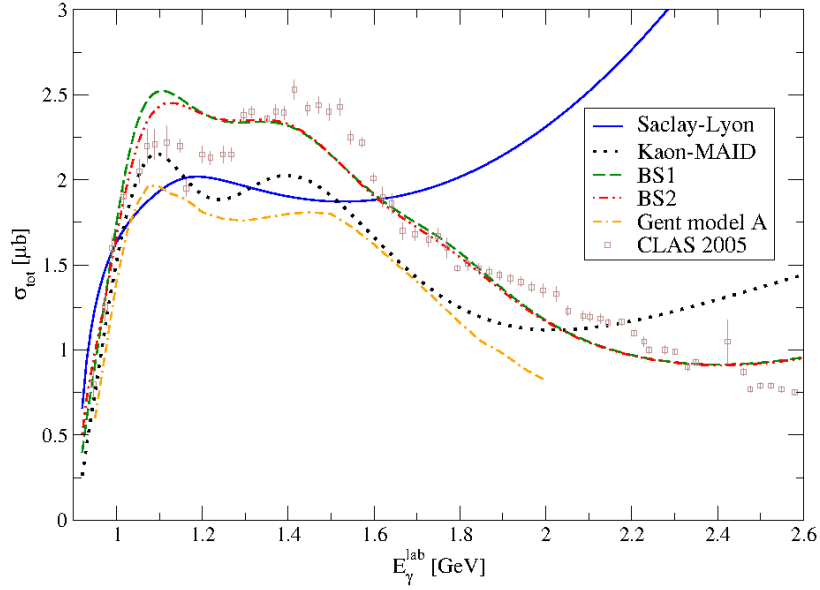}
\caption{(Color online) Model predictions of the $p(\gamma,K^+)\Lambda$ total cross section. For comparison, the Gent isobar model (model A) was added as read from Fig. 7 in Ref. \cite{Jan01A}. Notation of the rest of the curves is the same as in the Fig. \ref{fig:crs-theta}. Data stem from Fig. 20 in Ref.~\cite{CLAS05}.}
\label{fig:tot-crs}
\end{figure}

The new isobar models, eligible for the resonance region, describe 
data well up to energy $W\approx 2.4$~GeV. Above this energy the 
cross sections systematically rise overshooting the data, which is 
more apparent at forward angles in Fig.~\ref{fig:crs-w} and which is 
a well-known feature of isobar models. In the new models, the 
contributions of the nucleon resonances in the \emph{s} channel are 
regularized by the strong-enough hadron form factors as shown in 
Fig.~\ref{fig:hff}. The high-energy divergence is therefore created 
mainly by the background part of the amplitude. This divergent 
behavior, however, differs for various models: in the KM model, 
predictions start diverging at forward angles above 2.2 GeV 
(the maximum energy for which the model was constructed) but
predictions of the SL model strongly overshoot the data at backward 
angles above 2 GeV. This divergent behavior of the isobar models  
is also well seen in the energy dependence of the total cross 
section as shown in Fig.~\ref{fig:tot-crs}. Whereas the KM model 
begins to diverge at $E_{\gamma}^{lab}=2.2\,\mbox{GeV}$, 
\emph{i.e.}, beyond its scope, the SL model produces a divergent 
behaviour above $E_\gamma^{lab}=1.6\,\mbox{GeV}$. 
Note, however, that the KM, SL, and Gent models were fitted to 
the old SAPHIR data and, therefore, slightly underestimate 
the current CLAS data (see Fig. 20 in Ref. \cite{CLAS05}).

The spin-3/2 and spin-5/2 nucleon resonances contributing mainly 
in the central-angle region are also important in 
the forward-angle region. They contribute in combination 
with the background terms. Moreover, they give rather diverse 
results: the spin-3/2 resonances raise the cross section 
making the peak around $\theta_K^{c.m.}=45^\circ$ 
whereas the spin-5/2 resonances lead to a decrease of 
the cross section for kaon angles around $60^\circ$. 

\begin{figure}[b]
\includegraphics[width=11cm]{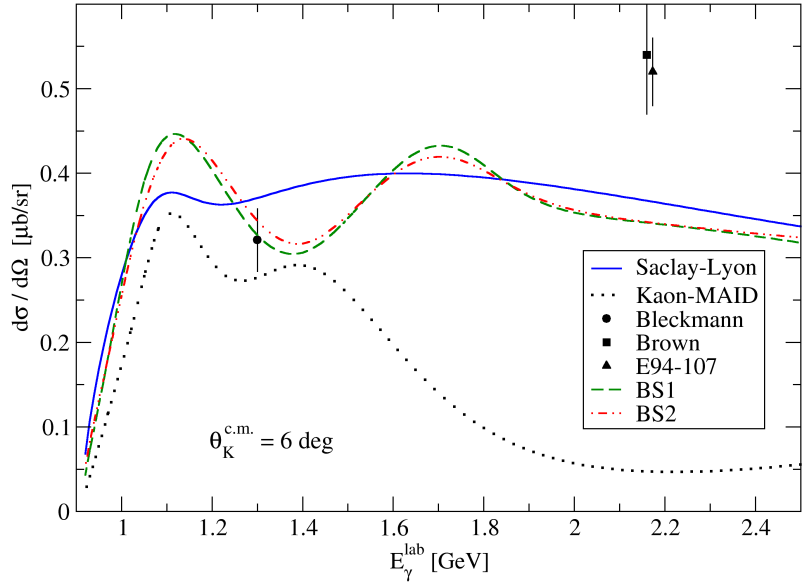}
\caption{(Color online) Predictions of the differential cross section for photoproduction 
at $\theta_K^{c.m.}=6^\circ$ is shown for several models. 
The only available photoproduction data point in this region 
is from Bleckmann \emph{et al}~\cite{Bleck}. The data points of
Brown~\cite{Brown} and E94-107~\cite{E94-107} are for 
electroproduction with a very small value of the virtual-photon 
mass $|k^2|$. Notation of the curves is the same as in the 
Fig. \ref{fig:crs-theta}.}
\label{fig:vfa}
\end{figure}
In the extreme forward-angle region, the discrepancies between different model 
predictions are substantial, especially for $E_{\gamma}^{lab}>1.5$~GeV, 
see Fig. \ref{fig:vfa}. The BS1, BS2, and Saclay-Lyon models predict similar 
magnitudes of the cross section in the whole energy range shown, but the Kaon-MAID 
model reveals a strong reduction of the results for higher energies due to 
suppression of the proton exchange by the hadron form factors. Recall that 
the BS1 and BS2 models also contain the form factors and that the strength 
they predict at small angles is made by another, more complex mechanism -- 
interference effects of the hyperon resonances with the Born terms and 
of higher-spin nucleon resonances with the background -- discussed above.
The energy dependence of the SL result is quite flat being dominated by 
the non-resonant proton exchange, which is not suppressed in SL, while the 
BS1 and BS2 models predict two broad peaks at 
$E_{\gamma}^{lab}=1.1\,\mbox{GeV}$ ($W=1.7$~GeV) and 
$E_{\gamma}^{lab}=1.7\,\mbox{GeV}$ ($W=2$~GeV). 
It is well-known that for kaon angles smaller than $\theta_K^{c.m.} = 25^\circ$ 
there are almost no available experimental data. Consequently, the models cannot 
be reliably tested in this region, which increases uncertainties in calculations of 
the hypernucleus production spectra~\cite{HProd, ByMa}.
In Fig.~\ref{fig:vfa}, 
the only data point for photoproduction is that by Bleckmann~\cite{Bleck} at 
$E_\gamma = 1.3$~GeV, which is consistent with all shown model predictions. 
The other two data points are from the measurements of electroproduction 
with almost real photons, {\em e.g.}, $-k^2=0.07$~(GeV/c)$^2$ for the JLab 
experiment E94-107~\cite{E94-107}, which prefer predictions of the SL, BS1, and 
BS2 models. 

\begin{figure}[t]
\includegraphics[width=8.5cm,angle=270]{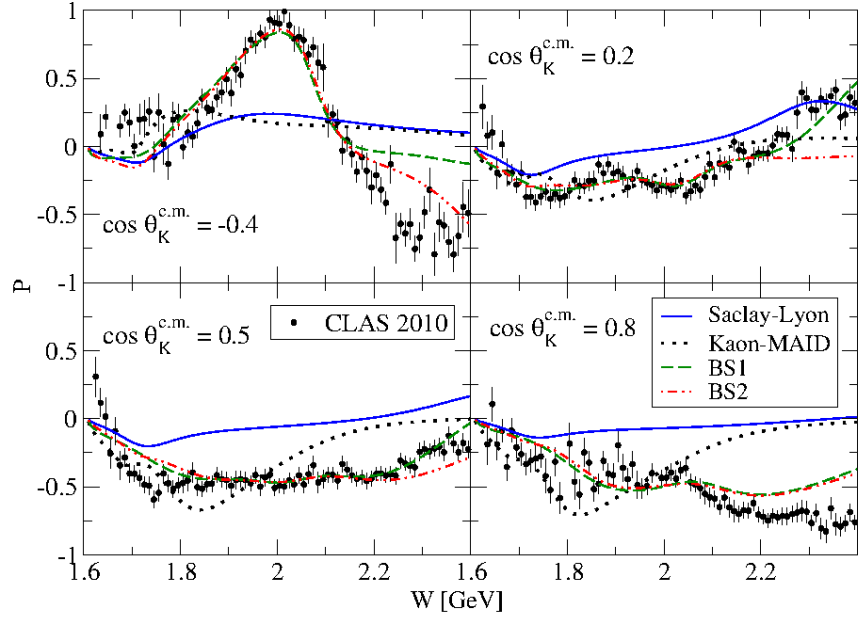}
\caption{(Color online) Results for energy dependent hyperon polarization 
are shown for several kaon angles 
$\theta_K^{c.m.}$. Notation of the curves is the same as in the 
Fig.~\ref{fig:crs-theta}. The CLAS data are from \cite{CLAS10}.}
\label{fig:pol-w}
\end{figure}

\begin{figure}[b]
\includegraphics[width=8.5cm,angle=270]{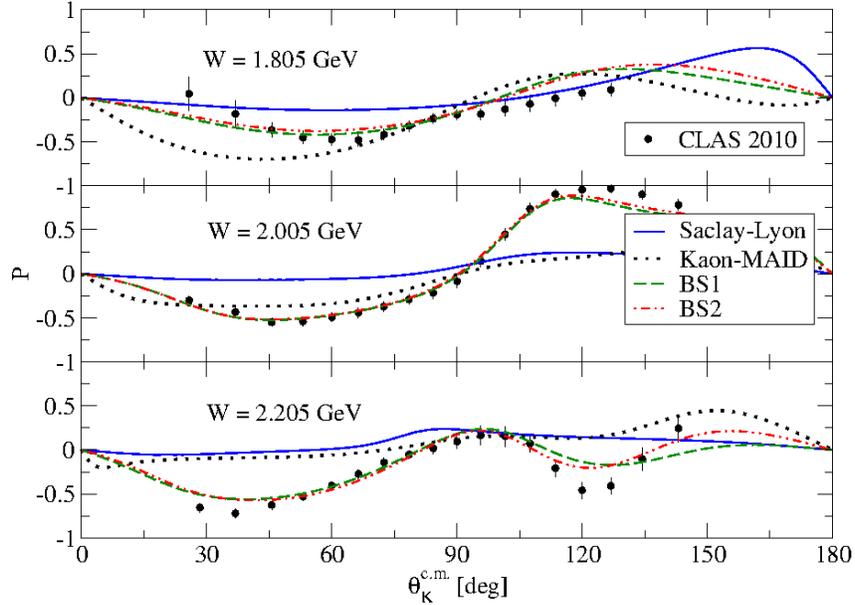}
\caption{(Color online) Results for the angular dependence of hyperon 
polarization are shown for several c.m. energies $W$. Notation of the curves 
is the same as in the Fig.~\ref{fig:crs-theta}. The CLAS data are from \cite{CLAS10}.}
\label{fig:pol-theta}
\end{figure}

\begin{figure}[t]
\includegraphics[width=9cm,angle=270]{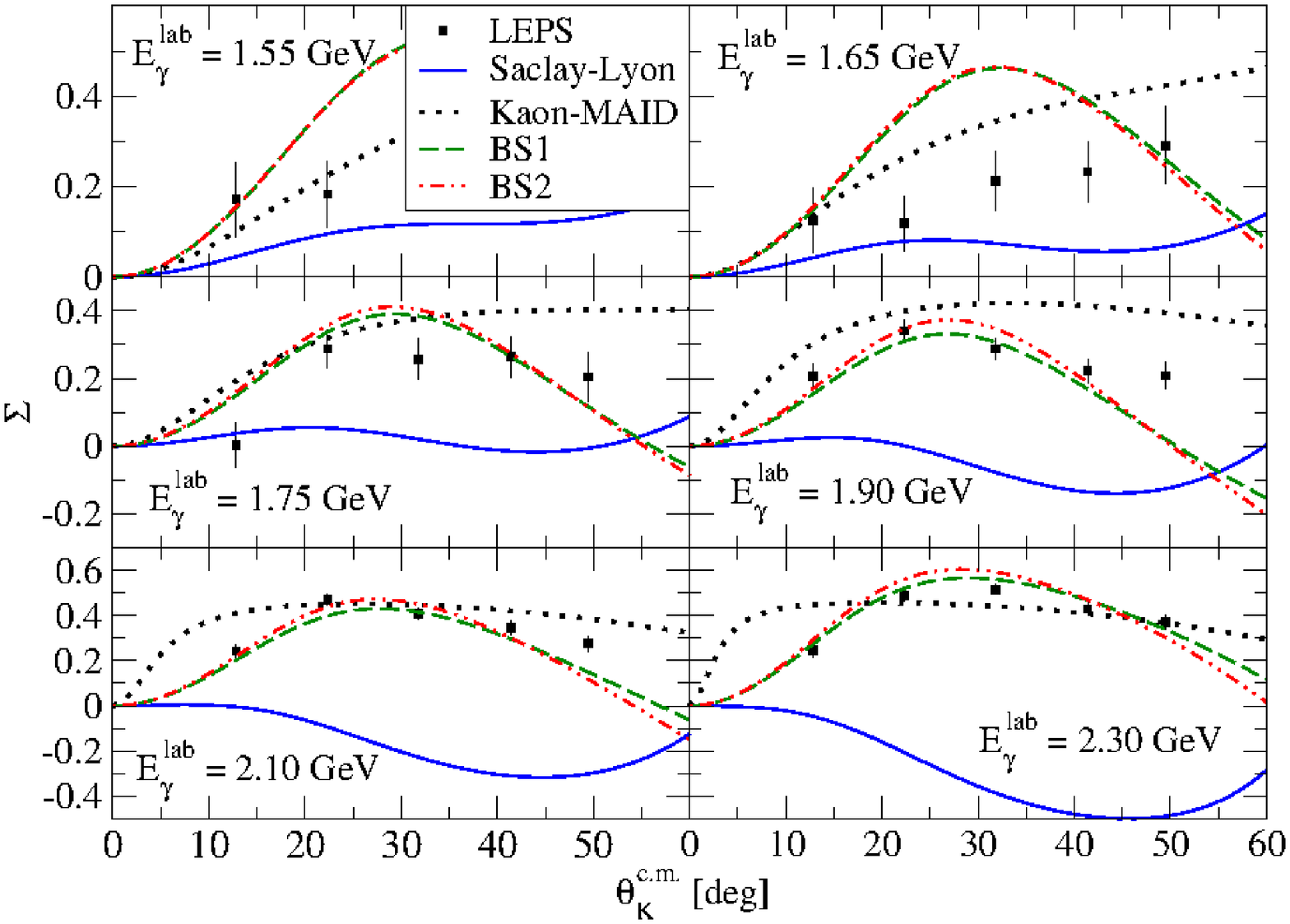}
\caption{(Color online) Results for the angular dependence of beam asymmetry are shown for several 
photon lab energies. Notation of the curves is the same as in the Fig. \ref{fig:crs-theta}. 
The LEPS data are from~\cite{LEPS}.}
\label{fig:sg-theta}
\end{figure}

The spin observables are very important in fine-tuning the interference among 
many different contributions. Plenty of new high-quality data for hyperon polarization 
and several tens for beam asymmetry and target polarization are now available. 
These data were also used in fitting the BS1 and BS2 models. 
In Figs.~\ref{fig:pol-w}, \ref{fig:pol-theta}, and \ref{fig:sg-theta}, we 
compare results of the models with the LEPS and CLAS data.

In the case of hyperon polarization, the Born terms on their own yield 
zero contribution but their interference with other terms appears 
to be important, especially the interference with the nucleon resonances.
The models were fitted to the hyperon polarization data from the threshold up to $2.225\,\mbox{GeV}$. In this energy range and mainly in the central-angle 
region, the data are captured by the BS1 and BS2 models well. 
On the other hand, the Saclay-Lyon and Kaon-MAID models do not even fit 
the shape of the data. Note, however, that these old models were not fitted 
to the hyperon polarization or beam-asymmetry data. 

For photon laboratory energy higher than $1.9\,\mbox{GeV}$, the BS1, BS2 and 
Kaon-MAID models describe the beam-asymmetry data satisfactorily, whereas 
the Saclay-Lyon model tends to underpredict the data in the whole energy 
range. Note that the data at lower energies, Fig.~\ref{fig:sg-theta}, 
have larger relative errors and therefore they cannot restrict the model 
parameters as much as the data for energies larger that 1.9~GeV.

\begin{figure}[t]
\includegraphics[width=10cm]{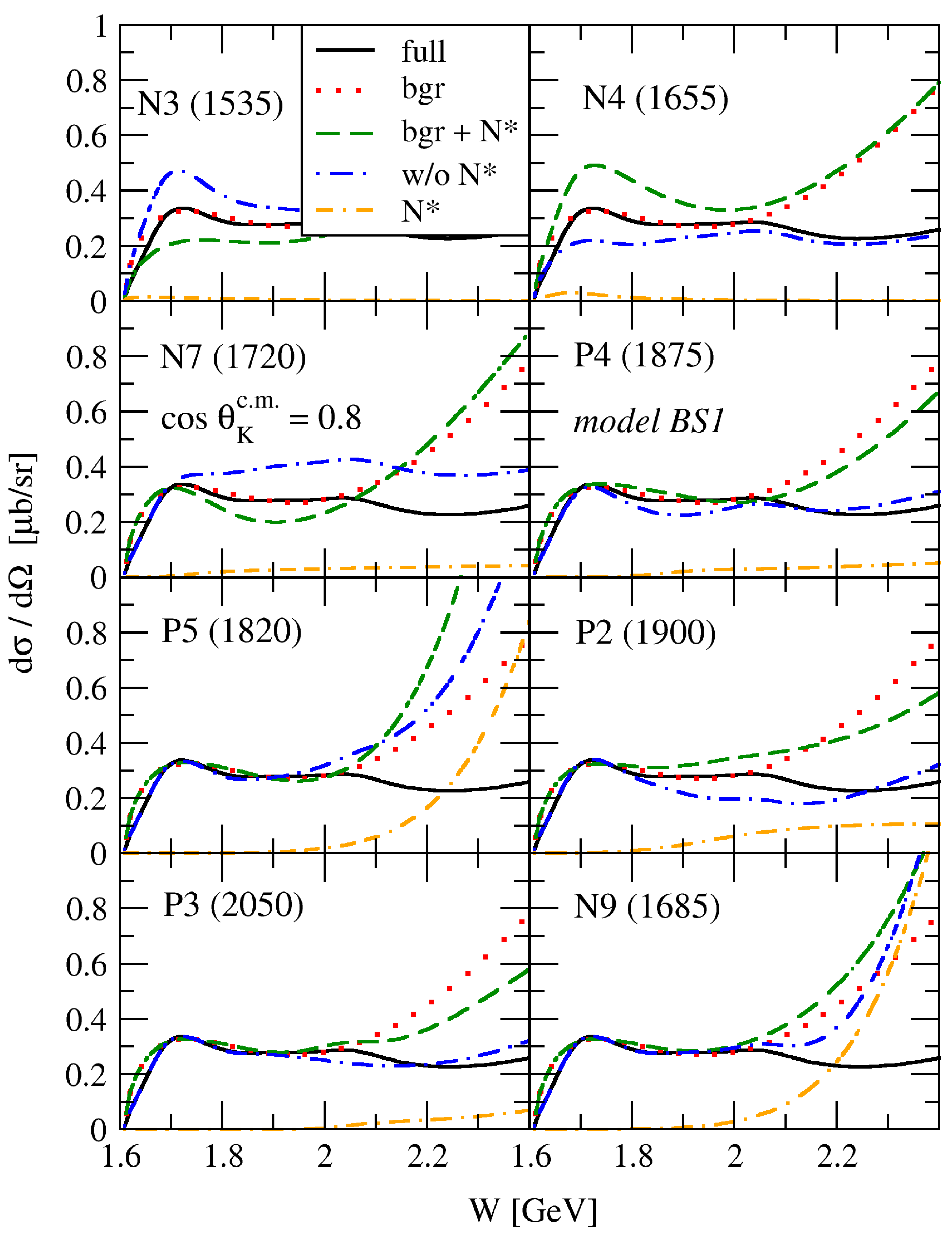}
\caption{(Color online) Double-polarization observables $C_x$ and $C_z$ are shown for various kaon angles. Since none of the models were fitted to the $C_x$ or $C_z$ data, the figure collects merely predictions of the models. Notation of the curves is the same as in the Fig. \ref{fig:crs-theta} and the data stem form the CLAS 2005 analysis~\cite{CLAS05}.}
\label{fig:N}
\end{figure}
The exchanges of the nucleon resonances in the $s$ channel constitute 
the resonant structure in the cross section. The effect of a particular 
resonance strongly depends on the magnitude and sign of its coupling constants, 
but this effect is hard to estimate in the kaon photoproduction due to overlapping  
of many resonances and occurrence of the complicated background. 
In Fig.~\ref{fig:N} we show effects of the nucleon resonances in the model BS1 
on the forward-angle differential cross section. A contribution of the resonance 
on its own, in its combination with the background, and a prediction of the full model 
without the resonance are shown. 
Comparing the latter with the full result one can infer an importance of 
the particular resonance in this kinematic region. 

In the BS1 model, the contribution of the subthreshold N3 resonance is small, 
as can be concluded from the relatively small value of its coupling 
parameter, Tab.~\ref{tab:Y(3/2)}. 
However, N3 significantly lowers, by 20 -- 30\%, the background contribution, 
which is important in the threshold region where it balances the contribution 
of N4. Omitting this resonance therefore leads to a growth of the cross section 
in the threshold region. 
Similarly, a strong effect is apparent for the N4, N7, and P2 resonances, where 
the latter two resonances affect the cross section rather at larger energies. 
On the other hand, the influence of the resonances P3, P5, and N9 on the forward-angle 
cross section is very small. Their influence is apparent only for energies above 2~GeV.  
The contributions of the spin-5/2 resonances N9 and P5 start to rise sharply 
around 2.2~GeV, which instigates the introduction of strong hadron form factors, 
\emph{e.g.}, the multidipole or multidipole Gauss~\cite{RPR11}. 
This effect is not seen for the P3 resonance because it is shifted to higher 
energies due to its larger mass. Since the BS2 model contains, except for the N6, 
the same nucleon resonances with very similar values of the coupling parameters, 
it behaves in a manner similar to the BS1 model. 
\begin{figure}[t]
\includegraphics[width=12cm]{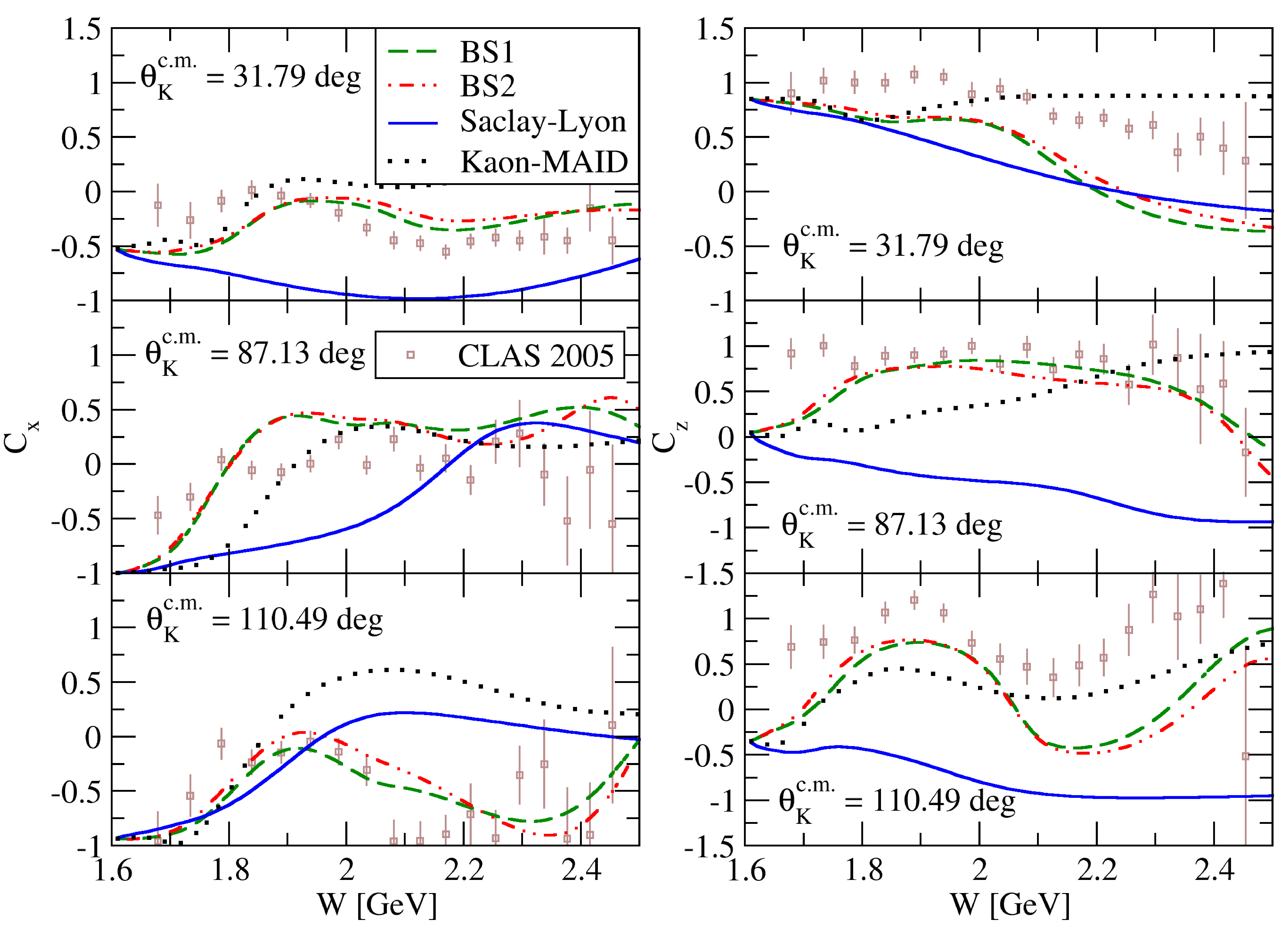}
\caption{(Color online) Model predictions of the $p(\gamma,K^+)\Lambda$ total cross section. For comparison, the Gent isobar model (model A) was added as read from Fig. 7 in Ref. \cite{Jan01A}. Notation of the rest of the curves is the same as in the Fig. \ref{fig:crs-theta}. Data stem from Fig. 20 in Ref.~\cite{CLAS05}.}
\label{fig:cxcz}
\end{figure}

In Fig. \ref{fig:cxcz}, the predictions of double-polarization observables $C_x$ and $C_z$ are shown for various kaon angles. Our new models as well as the well-known Kaon-MAID and Saclay-Lyon models were not fitted to these data sets. Therefore, the figure shows the predictive power of considered models. The Saclay-Lyon model fails to reproduce the $C_z$ data for larger kaon angles (whereas the data are positive, the model predictions have opposite sign). The correspondence between other model predictions and the data sets is considerably better: the Kaon-MAID predictions are of the same sign as the data and the BS models capture even the shape of the data.

Our findings on the nucleon resonances agree quite well with the results of 
the Bayesian analysis which used the Regge-plus-resonance model~\cite{RPR11}. 
In this analysis the N3, N4, and N7 resonances were assigned large relative 
probabilities, 13, 34, and 99, respectively, that they contribute to the kaon 
photoproduction process. Importance of these resonances was confirmed in our 
analysis. However, the resonances N9 and P3 were also shown to contribute 
significantly; their relative probabilities are 16 and 18, respectively, 
in the RPR-based analysis contrary to our findings which we attribute to the 
smaller energy window of our analysis (P3 and N9 contribute more at higher 
energies as shown in Fig.~\ref{fig:N}). 
In the Bayesian analysis, it was shown that the N5, N6, and N8 resonances 
are not required to describe the 
$\gamma\rm p \longrightarrow  K^+\Lambda$ data  which is also consistent 
with our conclusions, except for N6 in the BS2 model with the very small 
coupling parameter $G(N6)=-0.05$. The two-star spin-1/2 resonance $P_{11}(1880)$ 
(P1) was excluded in our analysis whereas it was included 
into the set of probable resonances in the Bayesian analysis with the relative 
probability 11.  The spin-5/2 state with near mass, $N^*(1860)$ (P5), was 
assumed in both new models instead. 
Note that adding P1 into the models does not improve the $\chi^2$ too much 
but it raises the number of considered resonances which we tried to keep as 
small as possible (according to the principle of the Occam's razor).
%
%
%
\section{Conclusion}
In this work, we have presented two new isobar models BS1 and BS2 for 
the $p(\gamma, K^+)\Lambda$ process in the energy range from threshold to 
$W=2.4$~GeV. The models provide satisfactory description of experimental 
data in the whole energy region and for all kaon angles. Their predictions 
for the cross sections at small kaon angles, being consistent with the results 
of the Saclay-Lyon model, suggest that the models can give reasonable values 
of the cross sections for the hypernucleus production. Construction of a new 
isobar model utilizing new precision data which could be used as an input 
in the hypernucleus calculations was one of the aims of this work.

In the construction of the single-channel models based on an effective 
Lagrangian we have utilized the consistent formalism by Pascalutsa for 
description of baryon fields with higher spin (3/2 and 5/2 in our case). 
This formalism ensures that only the physical degrees of freedom contribute 
in the baryon exchanges. Moreover, it provides regular amplitudes which 
are especially important for the $u$-channel exchanges allowing the inclusion 
of hyperon resonances with spin 3/2. These resonances were found to play 
an important role in description of the background part of the amplitude. 
They have not been considered in the older isobar models with the inconsistent 
formalism, except for version C of the Saclay-Lyon model~\cite{SLA}.

The set of selected nucleon resonances with spins 1/2, 3/2, and 5/2 
contributing most to the process agrees well with that selected in the Bayesian 
analysis with the Regge-plus-resonance model by the Gent group. We mostly confirm 
their result on the structure of the resonant part of the amplitude. 
The differences for the resonance part, \emph{e.g.}, different forms of 
the hadron form factor, stem from the fact that we limit our analysis only 
to the resonance region.
As for the missing resonances, we confirm importance of the $P_{13}(1900)$ 
and $D_{13}(1875)$ states for reasonable data description. We have found, 
however, that the spin-5/2 state $N^*(1860)$, recently included in the PDG 
Tables, is preferable to the spin-1/2 state $P_{11}(1880)$ included in 
the Bayesian analysis.  
 
Special attention was paid to the analysis of the background part of the 
amplitude which is important for a correct description of the forward-angle 
cross sections. In the background, which is a complicated effect of many 
various contributions in the isobar approach, the hyperon-resonance exchanges  
with spin 1/2 and 3/2  together with the Born terms appeared to be important 
components in the forward- and backward-angle regions, respectively. However, 
the current extensive data set still does not allow one to select the most 
significant hyperon resonances in the \emph{u} channel unambigously.

In the analysis, several forms of the hadron form factors were considered; 
we chose the dipole and multidipole forms as the most suitable for the data 
description. The obtained values of the cut-off parameters, around 2~GeV, 
suggest rather hard form factors.

The free parameters of the models were adjusted by fitting the cross 
section, hyperon polarization, and the beam asymmetry to new high-quality data 
from CLAS and LEPS and to older data. 
The overall number of resonances in the models, 15 and 16, is quite moderate 
in view of complexity of the kaon photoproduction in comparison with $\pi$ 
or $\eta$ photoproductions.

It is our desire and purpose to extend the model to study the electroproduction. 
The presented formalism can be easily extended in this line. 
Another possibility how to improve the model is to account for the unitarity 
by making the widths of the nucleon resonances energy-dependent functions 
as it was done, {\em e.g.}, in the Kaon-MAID model.   

\section*{ACKNOWLEDGEMENT}
The authors are very grateful to Bijan Saghai and Terry Mart for stimulating discussions and for providing the authors with the numerical codes of the Saclay-Lyon and Kaon-MAID models. This work was supported by the Grant Agency of the Czech Republic under Grant No. P203/15/04301.
%
%
%
\appendix
\section{Contributions to the invariant amplitude}
\label{sect:contrib}
We consider the process
\begin{equation}
\gamma_V(k)+p(p) \rightarrow K^+(p_K) + \Lambda (p_\Lambda)
\label{eq:process}
\end{equation}
with corresponding four-momenta given in the parentheses. The four-momentum of the intermediate particle is denoted by $q$. In the next sections, we summarize the invariant amplitudes with no hadron form factors. These are introduced in the manner shown in Appendix \ref{sect:contact}. The electromagnetic form factors are explicitly included in the Born contributions only. For the rest of the contributions, they are introduced merely by multiplying the coupling parameter with the appropriate electromagnetic form factor.

\subsection{Born \emph{s}-channel}
The electromagnetic vertex function reads
\begin{equation}
V^{EM}_{\mu}= F_{1}(k^2)\gamma_\mu+\frac{1-F_1(k^2)}{k^2}k_\mu \gamma\cdot k +\texttt{i} \frac{F_{2}(k^2)}{2m_p}\sigma_{\mu\nu}k^{\nu},
\end{equation}
where $F_{1}(k^2)$ and $F_{2}(k^2)$ are standard electromagnetic proton form factors, $F_1(0)=1$ and $F_2(0)=\kappa_p$, where $\kappa_p$ is anomalous proton magnetic moment. In the strong vertex, the pseudoscalar coupling is used 
\begin{equation}
V_{S} = \texttt{i}g_{K\Lambda p} \gamma_5.
\label{eq:PS-coupling}
\end{equation}

The invariant amplitude reads
\begin{equation}
\mathbb{M}_{Bs}=\bar{u}(p_{\Lambda})V_S\frac{\not \! p\,+\not \! k +m_{p}}{s-m_{p}^{2}}V_{\mu}^{EM}\varepsilon^{\mu}u(p),
\end{equation}

and can be cast into the form (\ref{eq:compact})
\begin{equation}
\begin{aligned}
\mathbb{M}_{Bs}=&\,\, \bar{u}(p_{\Lambda})\gamma_{5}\bigg[\mathcal{A}_{1}\mathcal{M}_{1}+\mathcal{A}_{2}\mathcal{M}_{2}+\mathcal{A}_{4}\mathcal{M}_{4} 
+ \mathcal{A}_{6}\mathcal{M}_{6}+g_{K\Lambda p}\frac{k\cdot \varepsilon}{k^{2}}\bigg]u(p),
\end{aligned}
\label{eq:Bs-gauge}
\end{equation}
where the last term in the brackets is the gauge-invariance breaking term. One then gets for the scalar amplitudes
\begin{subequations}
\begin{align}
\mathcal{A}_{1}&=\frac{g_{K\Lambda p}}{s-m_{p}^{2}}\left(F_1+F_2\right),\\
\mathcal{A}_{2}&=2\frac{g_{K\Lambda p}}{s-m_{p}^{2}}F_1\\
\mathcal{A}_{4}&=\frac{g_{K\Lambda p}}{s-m_{p}^{2}}
\frac{F_2}{m_{p}}=-2\mathcal{A}_{6}.
\end{align}
\end{subequations}

\subsection{Born \emph{t}-channel}
The electromagnetic vertex factor for pseudoscalar mesons $K^+$ reads
\begin{equation}
V_{\mu}^{EM}=F(k^2) (2p_K -k)_\mu+\frac{1-F(k^2)}{k^2}(2p_K -k)\cdot q \, q_\mu,
\end{equation}
where $F(0)=1$. The strong interaction vertex factor is the same as in (\ref{eq:PS-coupling}).
The invariant amplitude has the form
\begin{equation}
\mathbb{M}_{Bt}=\bar{u}(p_{\Lambda})V_S\frac{1}{t-m_{K}^{2}}V_{\mu}^{EM}\varepsilon^{\mu}u(p),
\end{equation}
which can again be cast to the compact form
\begin{equation}\mathbb{M}_{Bt}=\bar{u}(p_{\Lambda})\gamma_{5}\left[\mathcal{A}_{2}\mathcal{M}_{2}+\mathcal{A}_{3}\mathcal{M}_{3}-g_{K\Lambda p}\frac{k\cdot \varepsilon}{k^{2}}\right]u(p),
\label{eq:Bt-gauge}
\end{equation}
where the last term in the brackets is the same gauge-invariance breaking term as in the Born \emph{s}-channel contribution, Eq.~(\ref{eq:Bs-gauge}), but with the opposite sign. Therefore, these two terms cancel in the total amplitude of the process and the gauge invariance remains preserved.
There are only two nonzero scalar amplitudes
\begin{equation}
\mathcal{A}_{2}=-\mathcal{A}_3=2\frac{g_{K\Lambda p}}{t-m_{K}^{2}}F.
\end{equation}

\subsection{Born \emph{u}-channel}
The electromagnetic $\gamma \Lambda\Lambda$ vertex factor has the form
\begin{equation}
V_{\mu}^{EM} = F_1(k^2)\left[\gamma_\mu - \frac{k_\mu \gamma\cdot k}{k^2}\right]+\texttt{i}\frac{F_2(k^2)}{2m_\Lambda}\sigma_{\mu\nu}k^\nu,
\end{equation}
where $F_1(0)=0$ and $F_2(0)=\kappa_\Lambda$. The strong interaction vertex factor is the same as in (\ref{eq:PS-coupling}). The Born \emph{u}-channel amplitude reads
\begin{equation}
\mathbb{M}_{Bu}=\bar{u}(p_{\Lambda})V_{\mu}^{EM} \frac{\not\! p_{\Lambda} -\not \! k +m_{\Lambda}}{u-m_{\Lambda}^{2}}V_S \,\varepsilon^{\mu}u(p)
\end{equation}
and the scalar amplitudes $\mathcal{A}_j$ are
\begin{subequations}
\begin{align}
\mathcal{A}_{1}=&\frac{g_{K\Lambda p}}{u-m_{\Lambda}^{2}}(F_1+F_2),\\
\mathcal{A}_{3}=&2\frac{g_{K\Lambda p}}{u-m_\Lambda^2}F_1,\\
\mathcal{A}_{5}=&\frac{g_{K\Lambda p}}{u-m_{\Lambda}^{2}}
\frac{F_2}{m_{\Lambda}}= 2\mathcal{A}_{6}.
\end{align}
\end{subequations}

\subsection{Non-Born \emph{s}-channel: $N^{*}(1/2^\pm)$ exchange}
The amplitude for this contribution has the form
\begin{equation}
\begin{aligned}
\mathbb{M}^{N^*(1/2)}_{NBs}=&\,\, \texttt{i} \bar{u}(p_{\Lambda})\,g_{K \Lambda R}
\,\gamma_{5}\,\Gamma\frac{\not\! p + \not\! k +m_{R}}{s-m_{R}^{2}+\texttt{i}m_{R}\Gamma_{R}}  
\,\frac{\mu_{pR}}{m_{p}+m_R}\sigma^{\mu\nu}k_{\nu}\Gamma\varepsilon_{\mu}u(p).
\end{aligned}
\end{equation}
In the case of nucleon resonances we have to distinguish between the positive and negative parity resonances. This can be done by using $\Gamma$ in the form
\begin{equation}
\Gamma=
\begin{cases}
1, & P=+1,\\
\gamma_{5}, & P=-1,\\
\end{cases}
\label{eq:parityRel}
\end{equation}
The scalar amplitudes are
\begin{subequations}
\begin{align}
\mathcal{A}_{1}&=\frac{g_{K\Lambda R}}{s-m_{R}^{2}+\texttt{i}m_{R}\Gamma_{R}}\frac{m_R \pm m_p}{m_R + m_p}\mu_{pR},\\
\mathcal{A}_{4}&=\pm\frac{g_{K\Lambda R}}{s-m_{R}^{2}+\texttt{i}m_{R}\Gamma_{R}}\frac{2\mu_{pR}}{m_{p}+m_R}, \\
\mathcal{A}_{6}&=-\frac{1}{2}\mathcal{A}_4,
\end{align}
\end{subequations}
where the upper (lower) sign corresponds with the case of positive (negative) parity of the nucleon resonance.

\subsection{Non-Born \emph{s}-channel: $N^{*}(3/2^\pm)$ exchange}
The amplitude of the spin-3/2 contribution reads
\begin{equation}
\begin{aligned}
\mathbb{M}_{NBs}^{N^*(3/2)}=&\,\,\bar{u}(p_{\Lambda})\,\Gamma\, 
\frac{\texttt{i}f}{m_R m_K}\,\epsilon_{\mu\nu\lambda\rho}\,
\gamma_5\,\gamma^{\lambda} q^{\mu}p_{K}^{\rho} 
\,\frac{\not\! q + m_{R}}{s-m^{2}_{R}+\texttt{i}m_{R}\Gamma_{R}} 
\,\left(g^{\nu\beta}-\frac{1}{3}\gamma^{\nu}\gamma^{\beta}\right)\\
& \times \frac{1}{m_R(m_R+m_p)}\,\Big( g_1 q^\alpha F_{\alpha\beta} 
+ g_2 \not\! q\,F_{\beta\alpha}\,\gamma^\alpha 
- g_2 \gamma_\beta\,q^\alpha\,F_{\alpha\tau}\,\gamma^\tau\Big)\, 
\Gamma\,\gamma_{5}\,u(p),
\end{aligned}
\label{eq:N32ampl}
\end{equation}
where $g_1$ and $g_2$ are the electromagnetic coupling constants and $f$ is the strong coupling constant. Casting the amplitude to the compact form (\ref{eq:compact}), the individual scalar amplitudes $\mathcal{A}_{j}$ read
\begin{widetext}
\begin{subequations}
\begin{align}
\begin{split}
\mathcal{A}'_{1}={}&-\frac{G_{1}}{3}(q\cdot p_{\Lambda}\pm m_{R}m_{\Lambda})
\,q\cdot k+ \frac{G_{2}}{3}(2s \,q\cdot p_{\Lambda}-3s\,k\cdot p_{\Lambda}+
2s\,m_{p}m_{\Lambda}\mp m_{R}m_{\Lambda}\,q\cdot k\pm 2s\,m_{R}m_{\Lambda}\\
&\pm 2m_{p}m_{R}q\cdot p_{\Lambda} + 2q\cdot p_{\Lambda}\,q\cdot k),
\end{split}\\
\begin{split}
\mathcal{A}'_{2}={}&G_{1}\bigg[s\,k\cdot p_{\Lambda}\mp m_{R}m_{p}\,k \cdot p_{\Lambda}-\frac{1}{3}q\cdot p_{\Lambda}\,k^{2}\mp \frac{1}{3}m_{R}m_{\Lambda}k^{2}\bigg]+ G_{2}\left[-2s\, k\cdot p_{\Lambda}\mp\frac{1}{3}m_{\Lambda}m_{R}k^{2}+\frac{2}{3}k^{2}\,q\cdot p_{\Lambda}\right],
\end{split}\\
\begin{split}
\mathcal{A}'_{3} = {}& G_{1}(\pm m_{R}m_{p}-s)q\cdot k+G_{2}(2q\cdot k-k^{2})s,
\end{split}\\
\begin{split}
\mathcal{A}'_{4}={}&G_{1}\bigg[-\frac{1}{3}s\,m_{\Lambda}+\frac{1}{3}(m_{p}\mp m_{R})\,q\cdot p_{\Lambda}\pm \frac{1}{3}m_{\Lambda}m_{p}m_{R}\pm m_{R}\,k\cdot p_{\Lambda}\bigg]\\
& - G_{2}\left[-s\,m_{\Lambda}\mp \frac{1}{3}m_{\Lambda}m_{p}m_{R}+\frac{2}{3}m_{p}\,q\cdot p_{\Lambda}\right],
\end{split}\\
\begin{split}
\mathcal{A}'_{5} = {}&\mp G_{1}m_{R}\,q\cdot k + G_{2}(\pm m_{R}+m_{p})s,
\end{split}\\
\begin{split}
\mathcal{A}'_{6} = {}&G_{1}\bigg[\mp\frac{1}{3}m_{\Lambda}m_{p}m_{R}\mp m_{R}\,k\cdot p_{\Lambda}+\frac{1}{3}m_{\Lambda}s - \frac{1}{3}q\cdot p_{\Lambda}\,(m_{p}\mp m_{R})\bigg] \\ 
& + G_{2}\left[-\frac{1}{3}m_{\Lambda}s\mp\frac{1}{3}m_{\Lambda}m_{p}m_{R}+\frac{2}{3} q\cdot p_{\Lambda}\,(m_{p}\pm m_{R})\right],
\end{split}
\end{align}
\end{subequations}
\end{widetext}
where the coupling parameters $G_{1}$ and $G_{2}$ are given in Eq.~(\ref{eq:ccN32}) and the upper (lower) sign corresponds with the case of positive (negative) parity of the nucleon resonance.

Each amplitude $\mathcal{A}'_i, \, i=1,\ldots,6$, has to be multiplied by the propagator denominator
\begin{equation}
\mathcal{A}_i=\frac{1}{s-m_R^2+\texttt{i}m_R \Gamma_R} \mathcal{A}'_i.
\label{prop32}
\end{equation}

\subsection{Non-Born $s$-channel: $N^*(5/2^\pm)$ exchange}
The amplitude for the $N^*(5/2^\pm)$ exchange reads
\begin{equation}
\begin{aligned}
\mathbb{M}_{NBS}^{N^{*}(5/2)} = &-\frac{f}{m_K^4}\bar{u}(p_{\Lambda})\gamma_5 \Gamma \,q^2 p_K^\mu p_K^\nu \frac{\not \! q + m_R}{s-m_R^2+\texttt{i}m_R \Gamma_R}\\
& \times \mathcal{P}_{\mu\nu,\lambda\rho}(q)\,q^2\,p^\lambda \bigg[\frac{g_1}{(2m_p)^4}\,\gamma_\alpha\,F^{\alpha\rho}  
+ \frac{g_2}{(2m_p)^5} p_\alpha\,F^{\alpha\rho} \bigg]\Gamma\,u(p)
\end{aligned}
\label{eq:N52ampl}
\end{equation}
Casting the amplitude to the compact form (\ref{eq:compact}), the scalar amplitudes then read
\begin{widetext}

\begin{subequations}
\begin{align}
\begin{split}
\mathcal{A}'_1 = {}& G_1\bigg\{ \mp Q_{p_\Lambda p}Q_{kp_{\Lambda}}\pm \frac{1}{5}Q_{p_\Lambda p_\Lambda} Q_{kp}-\frac{1}{5} Q_{kp_\Lambda} (B\, q\cdot p + C m_p )+\frac{1}{5} Q_{p_\Lambda p} [2C m_p+(2s-q\cdot k)B] \bigg\}\\ 
& -\frac{G_2}{5}Q_{p_\Lambda p}C k\cdot p,
\end{split}\\
\begin{split}
\mathcal{A}'_2 = {}& G_1 \bigg\{ \pm Q_{p_\Lambda p} (k^2 q\cdot p_\Lambda - 2s k\cdot p_\Lambda)\mp \frac{1}{5}Q_{p_\Lambda p_\Lambda} k^2 (q\cdot p+s)-\frac{1}{5}[\mp 2q\cdot p_\Lambda\, k\cdot p_\Lambda\, q\cdot k s\pm k^2 (q\cdot p_\Lambda)^2 (q\cdot k+s)\\
&+2sm_R m_\Lambda k\cdot p_\Lambda q\cdot k-m_R m_\Lambda k^2 q\cdot p_\Lambda (q\cdot k+s)+C m_p k^2 q\cdot p_\Lambda]-\frac{1}{5}Q_{p_\Lambda p} k^2 B \bigg\}\\
&+G_2 \bigg\{ (m_R \pm m_p) Q_{p_\Lambda p} D-\frac{1}{5} (m_R \pm m_p)k^2 q\cdot p Q_{p_\Lambda p_\Lambda} \\
& +\frac{1}{5} (s k\cdot p_\Lambda-k^2 q\cdot p_\Lambda)(Bm_p q\cdot k -C k\cdot p)-\frac{1}{5}Q_{p_\Lambda p}B m_p k^2\bigg\},
\end{split}\\
\begin{split}
\mathcal{A}'_3 = {}& G_1 \bigg\{ \pm s Q_{p_\Lambda p} (2 k\cdot p+k^2)- \frac{1}{5}s [(2 k\cdot p\, q\cdot k -k^2 q\cdot p) B - m_p k^2 C ] \bigg\}\\
& + G_2 \bigg\{ s(m_R \pm m_p) k\cdot p \, Q_{p_\Lambda p}- \frac{1}{5} s k\cdot p ( B m_p q\cdot k- C k\cdot p ) \bigg\},
\end{split}\\
\begin{split}
\mathcal{A}'_4 = {}& G_1 \bigg\{ \frac{1}{5}(m_R \mp m_p)q\cdot k Q_{p_\Lambda p_\Lambda} -A Q_{p_\Lambda p}\\ 
& + \frac{1}{5}\{q\cdot p_\Lambda [B m_p q\cdot k+C (2k^2 + k\cdot p) +2s\,m_R k\cdot p_\Lambda]\pm 2 k\cdot p_\Lambda s^2 m_\Lambda \} \\
&-\frac{1}{5}Q_{p_\Lambda p} [m_\Lambda (m_R m_p \mp 3s)+(3m_R\mp m_p)q\cdot p_\Lambda] \bigg\}+G_2 \bigg\{\pm \frac{1}{5} k^2 q\cdot p Q_{p_\Lambda p_\Lambda}\mp D Q_{p_\Lambda p}+ \frac{1}{5}D E\\
&- \frac{1}{5} Q_{p_\Lambda p} [ m_R m_\Lambda (k^2\mp s)+q\cdot p_\Lambda(m_R m_p\mp k^2)\pm s(q\cdot p_\Lambda-m_\Lambda m_p) ]\bigg\},
\end{split}\\
\begin{split}
\mathcal{A}'_5 = {}& G_1 \bigg\{ s(\pm m_p-m_R)Q_{p_\Lambda p}- \frac{1}{5}s(B m_p q\cdot k - C k\cdot p ) \bigg\}+ G_2 \bigg\{ \pm s k\cdot p\, Q_{p_\Lambda p}+ \frac{1}{5}Es k\cdot p  \bigg\},
\end{split}\\
\begin{split}
\mathcal{A}'_6 = {}& G_1 \bigg\{ A Q_{p_\Lambda p} - \frac{1}{5} q\cdot p(\pm m_p -m_R)Q_{p_\Lambda p_\Lambda}-\frac{1}{5}q\cdot p_\Lambda ( B m_p q\cdot k-C k\cdot p)-\frac{1}{5}Q_{p_\Lambda p}[m_\Lambda (\pm s-m_R m_p)+A]\bigg\}\\
&-G_2 \bigg\{\pm q\cdot p_\Lambda \, k\cdot p\, Q_{p_\Lambda p} \pm \frac{1}{5} q\cdot p \,k\cdot p\, Q_{p_\Lambda p_\Lambda}+\frac{1}{5} q\cdot p_\Lambda \, k\cdot p E +\frac{1}{5} Q_{p_\Lambda p}B\, k\cdot p\bigg\},
\end{split}
\end{align}
\end{subequations}
\end{widetext} 
where the coupling parameters $G_1$ and $G_2$ are given as in (\ref{eq:ccN52}) and
\begin{subequations}
\begin{align}
A = {}&q\cdot p_\Lambda(\pm m_p - m_R), \\
B = {}&\pm q\cdot p_\Lambda - m_\Lambda m_R, \\
C = {}&\pm sm_\Lambda - m_R\, q\cdot p_\Lambda, \\
D = {}&k^2\, q\cdot p_\Lambda - s\, k\cdot p_\Lambda, \\
E = {}&m_p C - q\cdot p\, B,
\end{align}
\end{subequations}
where the upper (lower) sign corresponds with the case of positive (negative) parity of the nucleon resonance.

The terms $Q_{p_\Lambda p},\,Q_{kp_\Lambda},\,Q_{p_\Lambda p_\Lambda}$ and $Q_{kp}$ include four-momenta products given by the general prescription
\begin{equation}
Q_{XY} = s \, X\cdot Y - X\cdot q\, Y\cdot q,
\end{equation}
the notation of four-momenta is given in (\ref{eq:process}).

Each amplitude $\mathcal{A}'_i, \, i=1,\ldots,6$, has to be multiplied by 
the propagator denominator as in Eq.~(\ref{prop32}).

\subsection{Non-Born \emph{t}-channel: $K_1(1270)$ and $K^*(892)$ exchange}
The amplitude for the pseudovector meson $K_1(1270)$ ($J^{\pi}=1^{+}$) exchange reads
\begin{equation}
\begin{aligned}
\mathbb{M}_{NBt}^{K_1}=&\,\,\bar{u}(p_{\Lambda})\frac{g}{m}[g_{\alpha\mu}\,k\cdot (p-p_\Lambda)-k_\alpha (p-p_\Lambda)_\mu] 
\,\frac{\left(-g^{\alpha\lambda}+(p-p_{\Lambda})^{\alpha}(p-p_{\Lambda})^{\lambda}/m_{K_1}^{2}\right)}{t-m_{K_1}^{2}+\texttt{i}m_{K_1}\Gamma_{K_1}} \\
& \times \left[f_{V}\gamma_{\lambda}\gamma_{5}+\frac{f_{T}}{m_{\Lambda}+m_{p}}(\not\! p_{\Lambda}-\not\!p)\gamma_{\lambda}\gamma_{5}\right]\varepsilon^{\mu}u(p).
\end{aligned}
\end{equation}

And the scalar amplitudes $\mathcal{A}_j$ are given as
\begin{subequations}
\begin{align}
 \mathcal{A}_{2}&=\frac{-2G_{T}}{(m_\Lambda + m_p)(t-m_{K_1}^{2}+\texttt{i}m_{K_1}\Gamma_{K_1})} \,p_{\Lambda}\cdot k,\\
 \mathcal{A}_{3}&=\frac{2G_{T}}{(m_\Lambda + m_p)(t-m_{K_1}^{2}+\texttt{i}m_{K_1}\Gamma_{K_1})} \,p\cdot k,\\
 \mathcal{A}_{4}&=\frac{G_V+G_T (m_\Lambda - m_p)/(m_\Lambda + m_p)}{t-m_{K_1}^{2}+\texttt{i}m_{K_1}\Gamma_{K_1}},\\
 \mathcal{A}_{5}&= - \mathcal{A}_{4}.
\end{align}
\end{subequations}
with $G_{V,T} = g f_{V,T}/m$. The mass scale $m$ is arbitrarily chosen as $1\,\mbox{GeV}$.

\noindent
The vector meson $K^{*}(892)$ ($J^{\pi}=1^{-}$) exchange amplitude is
\begin{equation}
\begin{aligned}
\mathbb{M}_{NBt}^{K^*}=&\,\,\texttt{i}\bar{u}(p_{\Lambda})\frac{g}{m}\epsilon^{\mu\nu\alpha\beta}k_{\alpha}(p_{\Lambda}-p)_{\beta} 
\,\frac{\left(-g_{\nu\sigma}+(p-p_{\Lambda})_{\nu}(p-p_{\Lambda})_{\sigma}/m_{K^{*}}^{2}\right)}{t-m_{K^{*}}^{2}+\texttt{i}m_{K^{*}}\Gamma_{K^{*}}} \\
& \times \left[f_{V}\gamma^{\sigma}+\frac{f_{T}}{m_{\Lambda}+m_{p}}(\not\! p_{\Lambda}-\not\! p)\gamma^{\sigma}\right]\varepsilon_{\mu}u(p).
\end{aligned}
\end{equation}

The scalar amplitudes are given as
\begin{subequations}
\begin{align}
 \mathcal{A}_{1}&=\frac{G_V (m_\Lambda+m_p) + G_T \,t/(m_\Lambda+m_p)}{t-m_{K^{*}}^{2}+\texttt{i}m_{K^{*}}\Gamma_{K^{*}}},\\
 \mathcal{A}_{2}&=\frac{2 k\cdot p_\Lambda\, G_T}{(m_\Lambda+m_p)(t-m_{K^{*}}^{2}+\texttt{i}m_{K^{*}}\Gamma_{K^{*}})},\\
 \mathcal{A}_{3}&=\frac{-2 k\cdot p \, G_T}{(m_\Lambda+m_p)(t-m_{K^{*}}^{2}+\texttt{i}m_{K^{*}}\Gamma_{K^{*}})},\\
 \mathcal{A}_{4}&=\frac{G_{V} - G_T (m_\Lambda - m_p)/(m_\Lambda + m_p)}{t-m_{K^{*}}^{2}+\texttt{i}m_{K^{*}}\Gamma_{K^{*}}},\\
 \mathcal{A}_{5}&=\frac{G_{V} + G_T (m_\Lambda - m_p)/(m_\Lambda + m_p)}{t-m_{K^{*}}^{2}+\texttt{i}m_{K^{*}}\Gamma_{K^{*}}},
\end{align}
\end{subequations}
with $G_{V,T} = g f_{V,T}/m$. As in the pseudovector case, the mass $m$ is arbitrarily chosen to be $1\,\mbox{GeV}$.

\subsection{Non-Born \emph{u}-channel: $Y^{*}(1/2^{\pm})$ exchange}
The non-Born amplitude for the $Y^{*}(1/2^{\pm})$ exchange is
\begin{equation}
\begin{aligned}
\mathbb{M}_{NBu}^{Y^*(1/2)}=&\,\,\texttt{i} \bar{u}(p_{\Lambda})\frac{\kappa_{\Lambda R}}{m_{\Lambda}+m_{R}}\sigma^{\mu\nu}k_{\nu}\Gamma 
\,\frac{\not \! p_{\Lambda} - \not\! k + m_{R}}{u-m_{R}+\texttt{i}m_{R}\Gamma_{R}}g_{K\Lambda^{*}p}\gamma_{5}\Gamma\varepsilon_{\mu}u(p),
\end{aligned}
\end{equation}
with $\Gamma$ defined as in (\ref{eq:parityRel}).

The scalar amplitudes $\mathcal{A}_j$ are then
\begin{subequations}
\begin{align}
\mathcal{A}_{1}&=\frac{g_{K\Lambda^{*}p}}{u-m_{R}^{2}+\texttt{i}m_{R}\Gamma_{R}} \frac{m_{R}\pm m_{\Lambda}}{m_{R}+m_{\Lambda}}\kappa_{\Lambda R},\\
\mathcal{A}_{5}&=\pm\frac{g_{K\Lambda^{*}p}}{u-m_{R}^{2}+\texttt{i}m_{R}\Gamma_{R}}\frac{2\kappa_{\Lambda R}}{m_{\Lambda}+m_{R}},\\
\mathcal{A}_{6}&= \frac{1}{2}\mathcal{A}_{5},
\end{align}
\end{subequations}
where the upper (lower) sign corresponds with the positive (negative) parity of the resonance.

\subsection{Non-Born \emph{u}-channel: $Y^{*}(3/2^\pm)$ exchange}
The amplitude for the $Y^*(3/2^\pm)$ exchange in the \emph{u}-channel reads
\begin{equation}
\begin{aligned}
\mathbb{M}_{NBu}^{Y^*(3/2)} = &\,\,\,\bar{u}(p_\Lambda)\,\Gamma\,\gamma_5\, 
\frac{1}{m_R(m_R+m_\Lambda)}\Big[ g_1\,q^\alpha\,F_{\alpha\beta} 
+ g_2 \Big(\not\! q\,F_{\beta\alpha}\gamma^\alpha - \gamma_\beta\, 
q^\alpha\,F_{\alpha\rho}\,\gamma^\rho \Big) \Big] \\
& \times \frac{\not\!q + m_R}{u-m_R^2+\texttt{i}m_R\Gamma_R}\left(g^{\beta\nu}-\frac{1}{3}\gamma^\beta \gamma^\nu\right) 
\,\Gamma \frac{\texttt{i}f}{m_R m_K}\,\epsilon_{\mu\nu\lambda\rho}
\,\gamma_5 \gamma^\lambda q^\mu p_K^\rho\,u(p),
\label{eq:Y^*(3/2)ex}
\end{aligned}
\end{equation}
Casting the amplitude to the compact form, the scalar amplitudes are given as
\begin{widetext}
\begin{subequations}
\begin{align}
\begin{split}
\mathcal{A}'_1 = {}&  -\frac{1}{3}G_1 q\cdot k\,(\pm m_R m_p + q\cdot p) \\
& + \frac{1}{3}G_2 \left[\pm 5m_R m_p\, q\cdot k\pm 2m_Rm_p u + 2q\cdot p \, q\cdot k\pm 2 m_R m_\Lambda\, q\cdot p + 2u\,q\cdot p+2 m_\Lambda m_p u + 3 u\, p\cdot k\right],
\end{split}\\
\begin{split}
\mathcal{A}'_2 = {}&G_1 q\cdot k\,(\pm m_R m_\Lambda - u) + G_2 (2 q\cdot k\,u-uk^2\mp 4m_Rm_\Lambda\, q\cdot k),
\end{split}\\
\begin{split}
\mathcal{A}'_3 = {}&\,\, G_1\left\{ \frac{1}{3}k^2 (\pm m_p m_R + q\cdot p) + p\cdot k\,(u\mp m_R m_\Lambda)\right\}+ G_2 \left[\pm 4m_R m_\Lambda \,p\cdot k \mp \frac{5}{3} m_Rm_p k^2 - \frac{2}{3}q\cdot p\,k^2 - 2p\cdot k\,u \right],
\end{split}\\
\begin{split}
\mathcal{A}'_4 = {}&\mp G_1 m_R\,q\cdot k + G_2[\pm 4m_R\,q\cdot k+u(\pm m_R+m_\Lambda)],
\end{split}\\
\begin{split}
\mathcal{A}'_5 = {}&\,\, \frac{1}{3}G_1[q\cdot p\,(\pm m_R-m_\Lambda)\mp m_R m_p m_\Lambda + u\,m_p \pm 3m_R \,p\cdot k]\\
& + G_2 \left[\pm \frac{5}{3}m_R m_p m_\Lambda \mp 4m_R\,p\cdot k - 
\frac{1}{3}u\, m_p + \frac{2}{3}m_\Lambda \,q\cdot p \mp 
\frac{4}{3}m_R \,q\cdot p\right],
\end{split}\\
\begin{split}
\mathcal{A}'_6 = {}& \,\,\frac{1}{3}G_1[q\cdot p\,(\pm m_R-m_\Lambda)\mp m_R m_p m_\Lambda + u\,m_p \pm 3m_R\,p\cdot k]\\
& + G_2 \left[\pm \frac{5}{3}m_R m_p m_\Lambda \mp 4m_R\,p\cdot k \mp 2m_R\,q\cdot p - u\,m_p + \frac{2}{3}m_\Lambda\,q\cdot p \right],
\end{split}\\
\end{align}
\end{subequations}
\end{widetext}
where $G_{1,2}$ are given as in (\ref{eq:ccN32}) with $m_p$ replaced by $m_\Lambda$ and the upper (lower) sign corresponds with the case of positive (negative) parity of the hyperon resonance. 
Each amplitude $\mathcal{A}'_i, \, i=1,\ldots,6$, has to be multiplied by the propagator denominator
\begin{equation}
\mathcal{A}_i=\frac{1}{u-m_R^2+\texttt{i}m_R \Gamma_R} \mathcal{A}'_i.
\end{equation}

\section{Inclusion of hadron form factors and the gauge-invariance restoration}
\label{sect:contact}
The hadron form factors are included in a similar manner as the electromagnetic ones for a gauge-invariant vertex: it is sufficient to multiply the coupling parameter with the hadron form factor, $G \to FG$, where $G$ and $F$ are the coupling parameter and hadron form factor, respectively.

With the introduction of hadron form factors the gauge non-invariant terms in 
the \emph{s}- and \emph{t}-channel Born contributions no longer cancel each other 
and the gauge invariance is lost. In order to restore it, the contact term
\begin{equation}
\begin{aligned}
\mathbb{M}_{contact}=&-g_{K\Lambda p}\bar{u}_{\Lambda}(p_{\Lambda})\gamma_{5}\bigg[\frac{2p^{\mu}+\not\! k \gamma^{\mu}}{s-m_{p}^{2}}(\hat{F}_{DW}-F_{s})
+ \frac{2p_{K}^{\mu}}{t-m_{K}^{2}}(\hat{F}_{DW}-F_{t})\bigg]u_{p}(p)\varepsilon_{\mu},
\label{eq:contactTerm}
\end{aligned}
\end{equation}
is implemented. For $\hat{F}_{DW}$ the form
\begin{equation}
\hat{F}_{DW}=F_{s}(s)+F_{t}(t)-F_{s}(s)F_{t}(t),
\label{eq:FDW}
\end{equation}
introduced by Davidson and Workman~\cite{DW} is used. In the definition (\ref{eq:FDW}) it holds that $F_{s}(s=m_{p}^{2})=F_{t}(t=m_{K}^{2})=1$ and $\hat{F}_{DW}(s=m_{p}^{2},t)=\hat{F}_{DW}(s,t=m_{K}^{2})=1$, which prevents the poles in the contact-term contribution (\ref{eq:contactTerm}) from being reached.

%
%

\end{document}